\documentclass[10pt]{iopart}

\usepackage{iopams}
\usepackage[T1]{fontenc}
\usepackage[utf8]{inputenc}
\usepackage{textcomp}
\usepackage{float}
\usepackage{graphicx}
\usepackage{tabularx}
\usepackage{subfigure}
\usepackage{grffile}
\usepackage{caption}
\usepackage[dvipsnames]{xcolor}
\usepackage{ulem}
\usepackage{soul}
\usepackage{enumerate}
\usepackage{hyperref}
\usepackage{url}
\usepackage{cite}
\usepackage{comment}
\usepackage{todonotes}
\usepackage{multirow}
\usepackage{booktabs}
\usepackage{gensymb}

\graphicspath{{./Figures/}}

\setulcolor{blue}
\setstcolor{red}

\begin{document}

\title{3D particle-in-cell simulations of negative and positive streamers in C$_4$F$_7$N-CO$_2$ mixtures}

\author{Baohong Guo$^{1}$, Ute Ebert$^{1,2}$, Jannis Teunissen$^{1,*}$}

\address{$^1$ Centrum Wiskunde \& Informatica (CWI), Amsterdam, The Netherlands}
\address{$^2$ Department of Applied Physics, Eindhoven University of Technology, Eindhoven, The Netherlands}

\ead{jannis.teunissen@cwi.nl}

\vspace{10pt}

\begin{indented}
\item[]
\today
\end{indented}

\begin{abstract}
  We investigate negative and positive streamers in C$_4$F$_7$N-CO$_2$ mixtures through simulations.
  These mixtures are considered to be more environmentally friendly than the insulating gas SF$_6$ that is widely used in high voltage technology.
  Simulations are performed using a 3D particle-in-cell model.
  Negative streamers can propagate when the background field is close to the critical field.
  We relate this to their short conductive channels, due to rapid electron attachment, which limits their field enhancement.
  % \ue{As in air, there is no unique stability field, but thicker streamers can keep propagating in lower background electric fields.}
  Positive streamers also require a background field close to the critical field, and in addition a source of free electrons ahead of them.
  In our simulations these electrons are provided through an artificial stochastic background ionization process as no efficient photoionization process is known for these gases.
  In 3D, we can only simulate the early inception stage of positive discharges, due to the extremely high electric fields and electron densities that occur.
  Qualitative 2D Cartesian simulations show that the growth of these discharges is highly irregular, resulting from incoming negative streamers that connect to existing channels.
  The inclusion of a stochastic background ionization process also has an interesting effect on negative discharges: new streamers can be generated behind previous ones, thereby forming a chain of negative streamers.
\end{abstract}

% \vspace{2pc}
% \noindent{\it Keywords}: chemically active species, chemical production, energy efficiency, $G$-value, activation energy, pulsed streamer discharges
% \submitto{\PSST on August 17, 2023}
%\maketitle
\ioptwocol

\section{Introduction}\label{sec:intro}

\subsection{Eco-friendly alternative to the greenhouse gas SF$_6$}\label{sec:SF6-alternative}

% Sulfur hexafluoride (SF$_6$), an insulating gas, has been extensively used in electric power equipment, including gas circuit breakers (GCBs), gas-insulated switchgears (GISs) and gas-insulated transmission lines (GILs).
% It can efficiently suppress electric discharges due to its excellent properties for electrical insulation and current interruption.
Sulfur hexafluoride (SF$_6$), an insulating gas, has been widely used in electric power equipment such as gas circuit breakers and gas-insulated switchgear~\cite{christophorou1997}, due to its excellent properties for electrical insulation and current interruption~\cite{malik1978, boggs1990}. 
But it is also the most potent industrial greenhouse gas, with a global warming potential (GWP) of about 23500 times that of CO$_2$ over a 100-year horizon and an atmospheric lifetime of about 1000 years~\cite{ray2017, IPCC2021}.
Therefore there is an urgent need for exploring more sustainable alternatives.
The most promising candidate is perfluoronitrile (C$_4$F$_7$N), an electronegative gas developed by the 3M Company~\cite{kieffel2016}.
It has a relatively low GWP of about 1490, a relatively short atmospheric lifetime of about 22 years, good material compatibility with most electric power equipment, and a dielectric strength twice that of SF$_6$~\cite{rabie2018}.
Depending on the application, C$_4$F$_7$N is often mixed with buffer gases such as CO$_2$, N$_2$ or dry air, primarily due to its relatively high liquefaction temperature (e.g. $-4.7\,\degree$C at 0.1\,MPa), considerations regarding environmental sustainability and safety, as well as cost reduction and availability~\cite{nechmi2016}.
% However, due to the relatively high liquefaction temperature of -4.7\,$\degree$C at 0.1\,MPa, C$_4$F$_7$N is often mixed with buffer gases such as CO$_2$, N$_2$ or dry air for outdoor applications at low temperatures~\cite{nechmi2016}.

\subsection{Streamer discharges in C$_4$F$_7$N-CO$_2$ mixtures}\label{sec:streamers-in-eco-mixtures}

% The existing understanding of electrical insulation performance in air and SF$_6$ cannot easily be extended to new eco-friendly alternative gases.
Numerous theoretical and experimental investigations have been conducted to explore the electric breakdown and recovery properties of C$_4$F$_7$N mixtures for practical applications.
The decomposition pathways of C$_4$F$_7$N have been studied in~\cite{zhang2017a, fu2019, chen2019, chen2020a, li2020d, xiao2021a}.
% ~\cite{zhang2017a, yu2018, wu2018, fu2019, chen2019, zhang2019, chen2020, chen2020a, li2020d, ye2020, xiao2021a}.
Swarm parameters such as ionization and attachment coefficients have been experimentally measured to determine the reduced critical electric field and to estimate the dielectric strength of C$_4$F$_7$N mixtures~\cite{nechmi2017, chachereau2018, qin2019, yi2020, long2020, zhang2022}.
% ~\cite{nechmi2017, chachereau2018, hosl2019, qin2019, eguz2020, yi2020, long2020, zhang2022, zhang2023}.
Furthermore, their breakdown characteristics and dielectric properties have been extensively investigated through experimental measurements~\cite{tu2018, zhang2018, zhao2018, zhang2019a, nechmi2020, zhang2020b, zhou2020, nechmi2021, bahdad2022, lin2022} as well as theoretical calculations~\cite{zheng2019, zhang2020a, zheng2020}. 
% ~\cite{tu2018, zhang2018, zhao2018, li2019b, zhang2019a, zhang2019b, loizou2020, nechmi2020, zhang2020b, zhou2020, iddrissu2021, nechmi2021, bahdad2022, lin2022, meng2023}.
C$_4$F$_7$N-CO$_2$ mixtures with approximately 10\%--20\% C$_4$F$_7$N have been found to achieve a comparable dielectric strength to SF$_6$ under varying conditions.

% Many previous experimental studies have focused on measuring the dielectric strength of C$_4$F$_7$N mixtures under various conditions and comparing it to that of SF$_6$.
However, there are very few experimental studies on streamer discharges in C$_4$F$_7$N-CO$_2$ mixtures.
This is primarily due to the challenge of imaging such discharges, given the lack of detectable light emission.
Understanding streamer discharges in these mixtures is important, as they play a key role in the initial stage of electric breakdown~\cite{nijdam2020a}.
% This understanding can be achieved through quantitative modeling, which provides information on all particle densities and electric fields~\cite{guo2023}.
Several authors have therefore computationally studied streamers in C$_4$F$_7$N-CO$_2$ mixtures with 2D fluid simulations~\cite{vu-cong2020, wang2021a, fan2022, gao2022, yan2023, yan2023a}, which will be further elaborated in section~\ref{sec:comparison-past-work}.
The goal of this paper is to computationally study both negative and positive streamers in C$_4$F$_7$N-CO$_2$ mixtures in a full 3D geometry.
In contrast to previous simulations in these mixtures, we use a 3D particle model which can capture the stochastic nature and branching of streamers in a more realistic way.
% The thermal decomposition of C$_4$F$_7$N, which requires temperatures in the range of thousands of Kelvin~\cite{xiao2021a}, is not considered.
% Since there is no available data on the photoionization of C$_4$F$_7$N, we consider two sources of (initial) free electrons: a plasma seed near the electrode tip or stochastic background ionization.

\subsection{Cross sections for C$_4$F$_7$N}\label{sec:cross-sections-for-C4F7N}

% Electron impact ionization is one of the fundamental processes in streamer discharges.
A set of electron-neutral collision cross sections is required as the model input for accurate modeling of streamers in C$_4$F$_7$N-CO$_2$ mixtures~\cite{petrovic2009}.
% This cross section set generally consists of the momentum transfer cross section for elastic scattering and cross sections for inelastic scattering processes, including vibrational excitation, electronic excitation, impact ionization and electron attachment~\cite{zhang2023}.
In recent years, numerous studies have been dedicated to obtaining cross section data for C$_4$F$_7$N through both theoretical and experimental methods~\cite{zhang2020}.

The total electron impact ionization cross section of C$_4$F$_7$N was calculated using the Deutsch-M{\"a}rk (DM) formalism~\cite{xiong2017} and the binary-encounter-Bethe (BEB) method~\cite{wang2019a}.
To improve the agreement between these two methods, Zhong \textit{et al} proposed to combine the DM and BEB formalisms with a dual sigmoid function~\cite{zhong2019a}.
% combine the DM formalism at low electron energy and the BEB formalism at high electron energy, using a dual sigmoid function~\cite{zhong2019a}.
In~\cite{rankovic2019}, the total and partial ionization cross sections of C$_4$F$_7$N were theoretically and experimentally studied.
Furthermore, Chachereau \textit{et al} estimated the total electron attachment cross section of C$_4$F$_7$N by inversely calculating swarm parameters obtained from a pulsed Townsend experiment~\cite{chachereau2018}.
The dissociative electron attachment process was also experimentally examined in~\cite{rankovic2020}, but no cross section data were given.
In~\cite{sinha2020}, Sinha \textit{et al} utilized the spherical complex optical potential (SCOP) formalism to compute the total inelastic cross section of C$_4$F$_7$N, from which the electronic excitation and ionization cross sections were derived.
The SCOP method was further used to calculate the elastic cross section of C$_4$F$_7$N in~\cite{zhang2022b}.
Note that the above-mentioned cross sections were all considered individually.

A complete set of electron-neutral collision cross sections for C$_4$F$_7$N has recently been proposed for the first time in~\cite{zhang2023}, see figure~\ref{fig:cross_sections_C4F7N}, which will be used in the present paper.
Beyond electron-neutral collisions, the ion kinetics including detachment and ion conversion may also influence the dielectric properties of C$_4$F$_7$N-CO$_2$ mixtures, but there is currently only limited data available on these processes~\cite{hosl2019}.
% The authors began by constructing an initial cross section set for C$_4$F$_7$N through compilation and evaluation of the above individual cross sections. 
% Subsequently, they iteratively refined this initial set using the electron swarm method and validated its accuracy by systematically comparing swarm parameters obtained from the Boltzmann equation with experimental measurements.

\section{Simulation method}\label{sec:sim-method}

We simulate negative and positive streamers in C$_4$F$_7$N-CO$_2$ mixtures at 300\,K and 1\,bar.
Simulations are performed with a 2D/3D PIC-MCC (particle-in-cell, Monte Carlo collision) model, using the open-source \texttt{Afivo-pic} code~\cite{teunissen2016}.
Below we give a brief summary of the model, see more information in~\cite{teunissen2016, wang2022}.

\subsection{PIC-MCC model}\label{sec:pic-mcc}

\subsubsection{Particle mover and collisions}\label{sec:mover-collisions}

Electrons are tracked as particles. 
Ions are included as densities and assumed to be immobile on the considered short time scales (up to 50\,ns).
Neutral gas molecules are present as a homogeneous background that electrons stochastically collide with.

In the electrostatic approximation, the positions $\boldsymbol{\mathrm x}$ and velocities $\boldsymbol{\mathrm v}$ of simulation particles are advanced in time with the `Velocity Verlet' scheme~\cite{verlet1967} as
\begin{equation}
\label{eq:x-advance}
    \boldsymbol{\mathrm x}(t + \Delta t) = \boldsymbol{\mathrm x}(t) + \Delta t \, \boldsymbol{\mathrm v}(t) + \frac{1}{2} (\Delta t)^2 \, \boldsymbol{\mathrm a}(t)\,,
\end{equation}
\begin{equation}
\label{eq:v-advance}
    \boldsymbol{\mathrm v}(t + \Delta t) = \boldsymbol{\mathrm v}(t) + \frac{1}{2} \Delta t \, [\boldsymbol{\mathrm a}(t) + \boldsymbol{\mathrm a}(t + \Delta t)]\,,
\end{equation}
where $\boldsymbol{\mathrm a} = - (e/m_\mathrm{e}) \, \boldsymbol{\mathrm E}$ is the acceleration due to the electric field $\boldsymbol{\mathrm E}$, $e$ the elementary charge, and $m_\mathrm{e}$ is the electron mass.

Electron-neutral collisions are treated using the null-collision method~\cite{koura1986} assuming isotropic scattering, with collision rates derived from cross sections, see section~\ref{sec:cross-sections}.
We only consider electron-neutral collisions since the ionization degree of the discharges is typically below $10^{-4}$.

\subsubsection{Super-particles}\label{sec:super-particles}

In the PIC-MCC model, the stochastic development of discharges can be effectively captured by tracking individual electrons' trajectories and interactions.
However, it is computationally infeasible to simulate every electron individually due to the large number of electrons (above $10^8$) in a typical discharge.
To address this, super-particles are used, with their weights $w$ ($w \geqslant 1$) determining the number of physical electrons they represent.
Such super-particles can be merged or split between time steps using adaptive particle management~\cite{teunissen2014a}, thereby adjusting their weights $w$ to a desired weight $w_\mathrm{d}$ as
\begin{equation}
\label{eq:desired-weight}
    w_\mathrm{d} = n_\mathrm{e} \, \Delta V / N_\mathrm{ppc}\,,
\end{equation}
where $n_\mathrm{e}$ is the electron density in a grid cell, $\Delta V$ the cell volume, and $N_\mathrm{ppc}$ is the desired number of particles per cell, which is here set to $N_\mathrm{ppc} = 100$.

Therefore, super-particles at low electron densities represent few (or even single) physical electrons, whereas those at high electron densities represent multiple physical electrons.

\subsubsection{Adaptive mesh refinement for the electric field}\label{sec:AMR-electric-field}

Simulation particles are mapped to grid densities using a standard bilinear (2D) or trilinear (3D) weighting scheme.
Near refinement boundaries, the mapping is locally switched to the `nearest grid point’ scheme, to maintain the conservation of particle densities~\cite{wang2022}.

Subsequently, the electric field $\boldsymbol{\mathrm E}$ is calculated as $\boldsymbol{\mathrm E} = - \nabla \phi$. 
The electric potential $\phi$ is obtained by solving Poisson's equation
\begin{equation}
\label{eq:Poisson-equation}
    \nabla^2 \phi = - \rho/\varepsilon_0\,,
\end{equation}
where $\rho$ is the space charge density and $\varepsilon_0$ is the vacuum permittivity.
Equation (\ref{eq:Poisson-equation}) is solved using the geometric multigrid method included in the Afivo library~\cite{teunissen2018, teunissen2023}.
The calculated electric field is then interpolated from the grid back to particles using standard bilinear or trilinear interpolation.

For computational efficiency, the \texttt{Afivo-pic} code includes adaptive mesh refinement.
The mesh is refined if
% determined by the refinement criterion
% \begin{equation}
% \label{eq:refinement-criterion}
%     \alpha_\mathrm{eff}(E) \Delta x  > c_0\,,
% \end{equation}
$$\alpha_\mathrm{eff}(E) \Delta x  > 1\,,$$
where $\Delta x$ is the grid spacing and $\alpha_\mathrm{eff}(E)=\alpha(E)-\eta(E)$ is the field-dependent effective ionization coefficient, see figure~\ref{fig:transport_data_comparison}.
The minimal grid spacing is less than 1\,$\mu$m in the simulations.
Note that we here use the effective ionization coefficient $\alpha_\mathrm{eff}(E)$ instead of the ionization coefficient $\alpha(E)$ utilized previously in~\cite{teunissen2016, wang2022}.
% This is due to the relatively high electric field present in streamers in C$_4$F$_7$N-CO$_2$ mixtures, see e.g. section~\ref{sec:pos-3d-results}.

\subsection{Input data}\label{sec:input-data}

\subsubsection{Cross sections}\label{sec:cross-sections}

\begin{figure}
    \centering
    \includegraphics[width=0.48\textwidth]{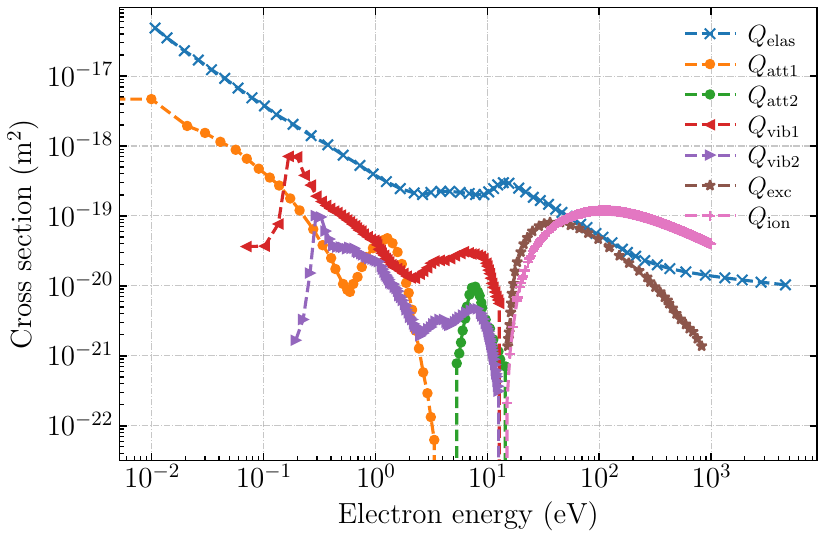}
    \caption{A complete set of electron-neutral collision cross sections for C$_4$F$_7$N from~\cite{zhang2023}. 
    This set includes elastic cross section ($Q_\mathrm{elas}$), electron attachment cross sections ($Q_\mathrm{att1}$, $Q_\mathrm{att2}$), vibrational excitation cross sections ($Q_\mathrm{vib1}$, $Q_\mathrm{vib2}$), electronic excitation cross section ($Q_\mathrm{exc}$) and electron impact ionization cross section ($Q_\mathrm{ion}$).}
    \label{fig:cross_sections_C4F7N}
\end{figure}

We use cross sections for elastic, vibrational excitation, electronic excitation, ionization and attachment collisions for C$_4$F$_7$N from the XJTUAETLab database~\cite{XJTUAETLab_database, zhang2023}, as shown in figure~\ref{fig:cross_sections_C4F7N}.
This database provides a complete cross section set for C$_4$F$_7$N.

Electron-neutral collision cross sections for CO$_2$ were obtained from the IST-Lisbon database~\cite{IST-Lisbon_database}.
This database includes an effective momentum-transfer cross section, considering both elastic and inelastic processes~\cite{grofulovic2016}.
For PIC simulations, an estimated elastic cross section was obtained by subtracting the total inelastic cross sections.
% To derive an estimated elastic cross section that can be used in PIC simulations, the total inelastic cross sections were subtracted.

For comparison of electron transport data in different gases, electron-neutral collision cross sections for air were taken from the Phelps database~\cite{eco_phelps_database}.

\subsubsection{Electron transport data}\label{sec:transport-data}

\begin{figure*}
    \centering
    \includegraphics[width=1\textwidth]{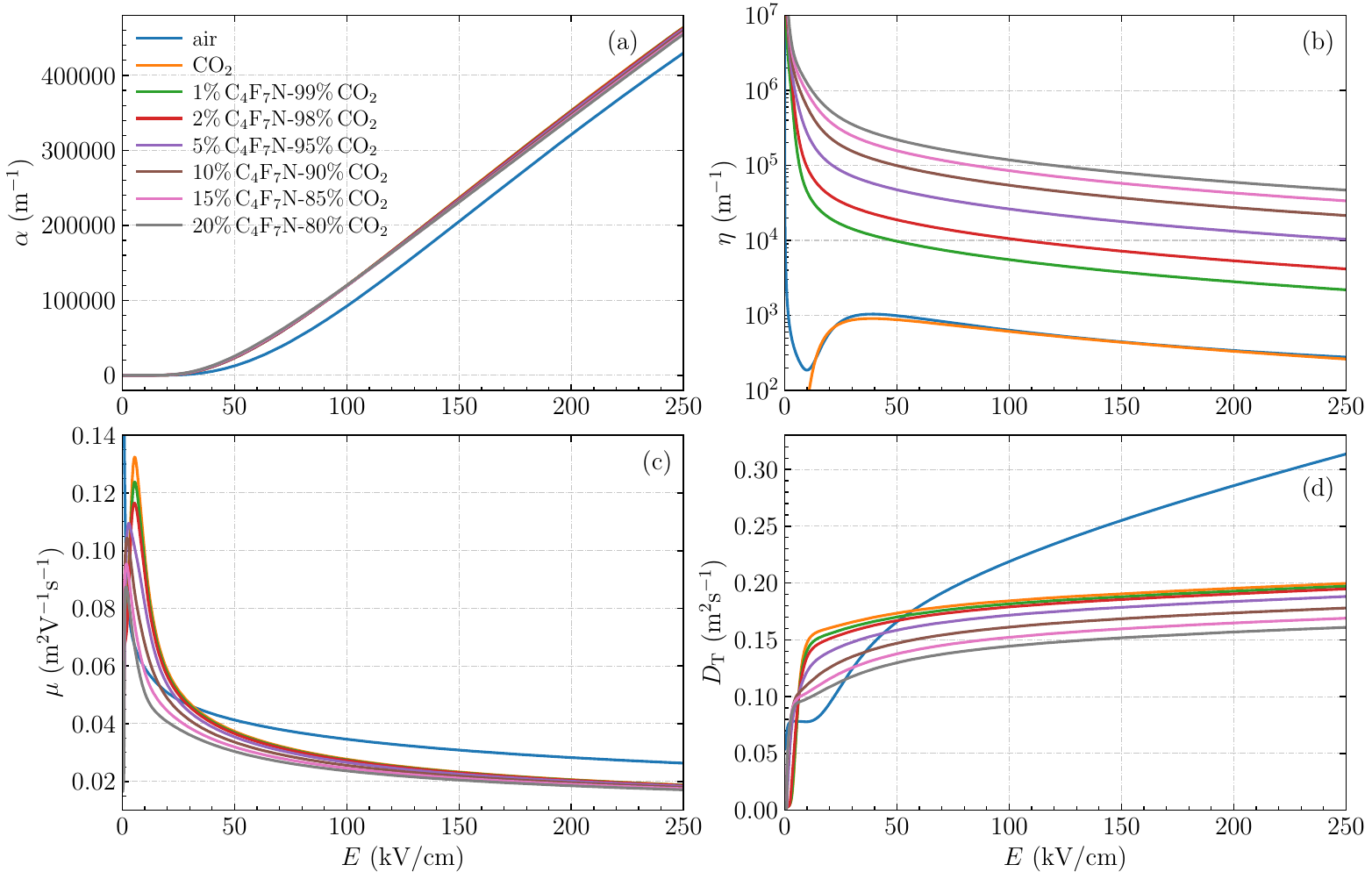}
    \caption{Comparison of electron transport data in air, pure CO$_2$ and C$_4$F$_7$N-CO$_2$ mixtures.
    (a) Ionization coefficient $\alpha$, (b) attachment coefficient $\eta$, (c) flux mobility $\mu$ and (d) flux transverse diffusion coefficient $D_\mathrm{T}$.
    These coefficients were computed with BOLSIG$+$~\cite{hagelaar2005} at 300\,K and 1\,bar, with cross sections for artificial dry air (80\% $\mathrm N_2$ and 20\% $\mathrm O_2$) obtained from the Phelps database~\cite{eco_phelps_database}, cross sections for CO$_2$ obtained from the IST-Lisbon database~\cite{IST-Lisbon_database}, and cross sections for C$_4$F$_7$N obtained from the XJTUAETLab database~\cite{XJTUAETLab_database}.}
    \label{fig:transport_data_comparison}
\end{figure*}

Differences in electron transport data for various C$_4$F$_7$N-CO$_2$ mixtures, pure CO$_2$ and air are illustrated in figure~\ref{fig:transport_data_comparison}.
These transport coefficients were computed with BOLSIG$+$~\cite{hagelaar2005} at 300\,K and 1\,bar, using cross sections described in section~\ref{sec:cross-sections}.
We remark that in the PIC-MCC model, these transport coefficients are not used since electron-neutral collisions are directly determined by cross sections.

% Note that in high electric fields, e.g., ahead of positive streamers in C$_4$F$_7$N-CO$_2$ mixtures (see section~\ref{sec:pos-3d-results}), the two-term approximation for electron transport data can cause errors of about 1\%--10\%.
Figure~\ref{fig:transport_data_comparison} shows that the ionization coefficient $\alpha$, flux mobility $\mu$ and flux transverse diffusion coefficient $D_\mathrm{T}$ are similar between these different gases.
However, the attachment coefficients $\eta$ in C$_4$F$_7$N-CO$_2$ mixtures are several orders of magnitude higher than those in pure CO$_2$ and air, leading to higher critical electric fields $E_\mathrm{k}$ (where the ionization coefficient $\alpha$ is equal to the attachment coefficient $\mu$), as summarized in table~\ref{tab:critical-field}. 

\begin{table}
\centering
% \captionsetup{width=0.5\textwidth}
\caption{The critical electric field $E_\mathrm{k}$ for different gases at 300\,K and 1\,bar, corresponding to figure~\ref{fig:transport_data_comparison}.}
\label{tab:critical-field}
\begin{tabular*}{0.48\textwidth}{c@{\extracolsep{\fill}}c}
  \br
  Gas & Critical electric field $E_\mathrm{k}$ \\
  \mr
  air & 28.2\,kV/cm \\
  CO$_2$ & 22.5\,kV/cm \\
  1\% C$_4$F$_7$N-99\% CO$_2$ & 40.2\,kV/cm \\
  2\% C$_4$F$_7$N-98\% CO$_2$ & 47.2\,kV/cm \\
  5\% C$_4$F$_7$N-95\% CO$_2$ & 61.0\,kV/cm \\
  10\% C$_4$F$_7$N-90\% CO$_2$ & 76.4\,kV/cm \\
  15\% C$_4$F$_7$N-85\% CO$_2$ & 88.6\,kV/cm \\
  20\% C$_4$F$_7$N-80\% CO$_2$ & 99.7\,kV/cm \\
  \br
\end{tabular*}
\end{table}

% Note that of the coefficients presented in figure~\ref{fig:transport_data_comparison} are only used to control the adaptive mesh refinement and for comparison.
% In the PIC-MCC model, electron-neutral collisions are directly determined by cross sections.

\subsection{3D computational domain}\label{sec:comput-domain-3d}

\begin{figure*}
    \centering
    \includegraphics[width=0.82\textwidth]{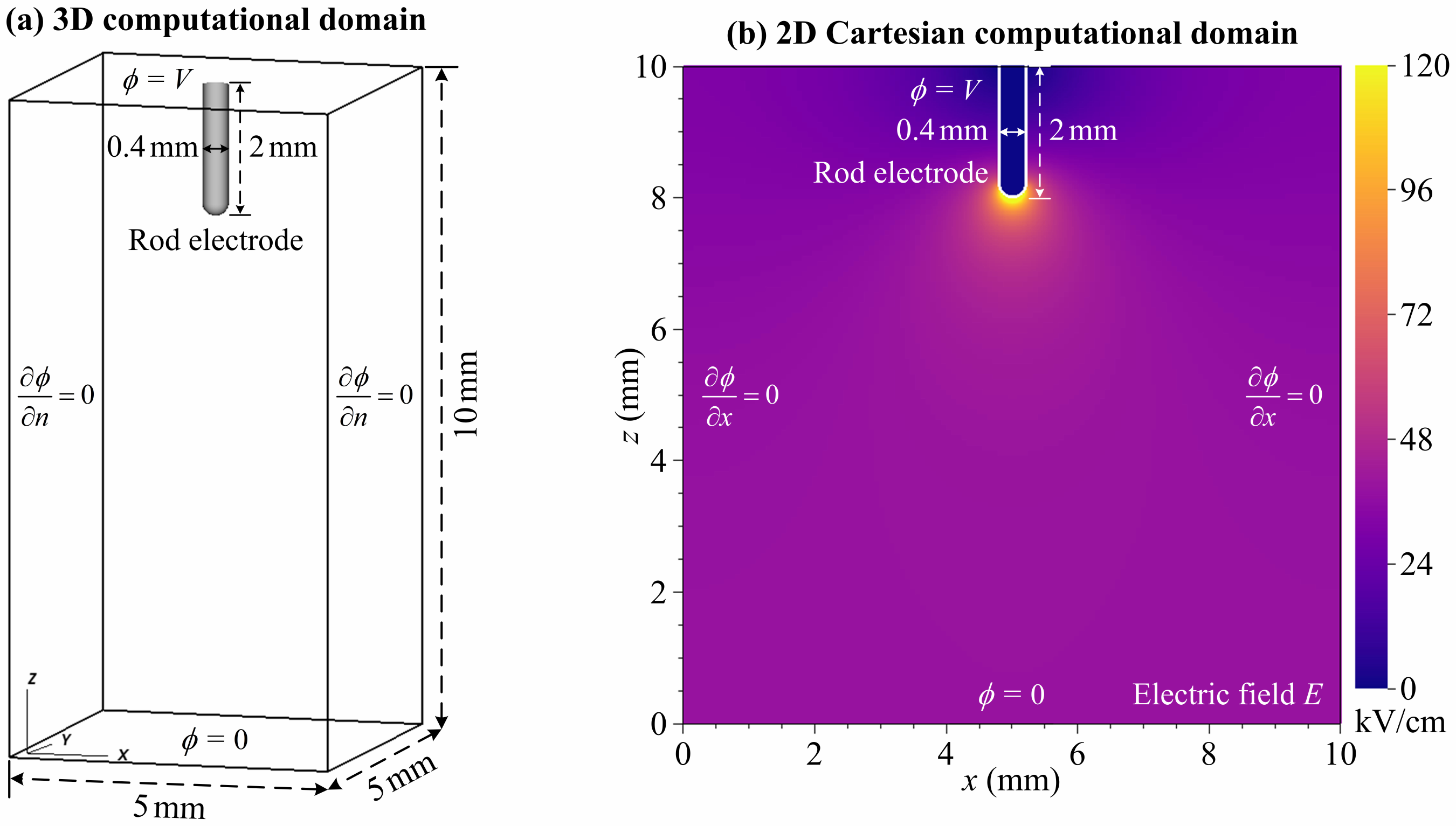}
    \caption{A view of (a) the 3D (5\,mm\,$\times$\,5\,mm\,$\times$\,10\,mm) and (b) the 2D (10\,mm)$^2$ Cartesian computational domains.
    Panel (b) also shows the electric field profile $E$ without a discharge at an applied voltage $V = - 36$\,kV. 
    % in a background field of $E_\mathrm{bg}=36$\,kV/cm.
    The centrally positioned rod electrode, protruding from the upper plate, has a length of 2\,mm and a diameter of 0.4\,mm.
    Boundary conditions for the electric potential $\phi$ are indicated.}
    \label{fig:computational_domain}
\end{figure*}

We simulate negative and positive streamers in C$_4$F$_7$N-CO$_2$ mixtures at 300\,K and 1\,bar using the 3D computational domain illustrated in figure~\ref{fig:computational_domain}(a), which measures 5\,mm\,$\times$\,5\,mm\,$\times$\,10\,mm.
The domain has a plate-plate geometry with a centrally positioned electrode protruding from the upper plate.
The electrode is rod-shaped with a semi-spherical tip, measuring 2\,mm in length and 0.4\,mm in diameter.

For the electric potential, a constant high voltage $V$ is applied on the upper plate and the rod electrode, the lower plate is grounded, and homogeneous Neumann boundary conditions are applied on the other sides of the domain.
Simulation particles are removed from the simulation if they enter the electrode or move beyond the domain boundaries.
Secondary electron emission from the electrode is not included.

\begin{figure}
    \centering
    \includegraphics[width=0.48\textwidth]{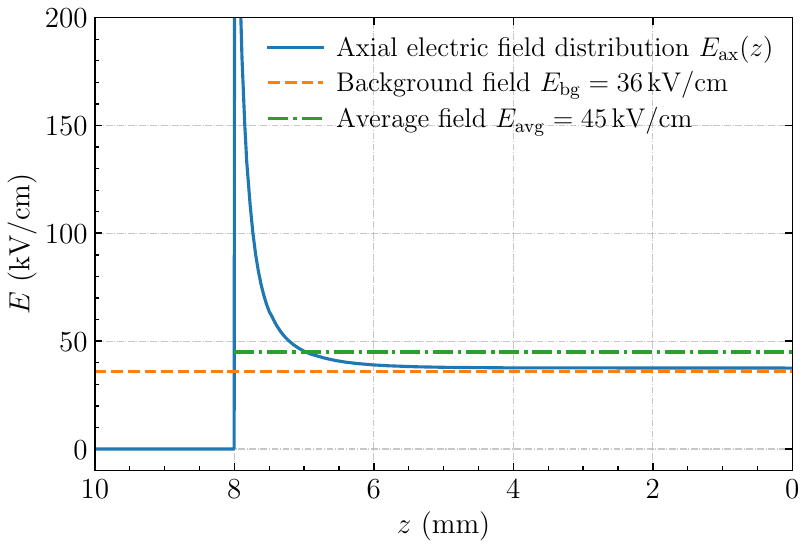}
    \caption{Axial electric field distribution $E_\mathrm{ax}(z)$ without a discharge for an applied voltage $V = - 36$\,kV.
    The background electric field $E_\mathrm{bg}$ and the average electric field $E_\mathrm{avg}$ are also indicated.}
    \label{fig:background_field}
\end{figure}

The axial electric field $E_\mathrm{ax}(z)$ away from the rod electrode relaxes to the average field between the two plate electrodes, see figure~\ref{fig:background_field}.
We therefore define the background electric field $E_\mathrm{bg}$ as
\begin{equation}
  \label{eq:bg-field}
  E_\mathrm{bg} = |V| / d_\mathrm{plates}\,,
\end{equation}
where $d_\mathrm{plates} = 10$\,mm is the distance between the two plate electrodes.
Note that $E_\mathrm{ax}(z) \approx E_\mathrm{bg}$ in most of the domain.
In addition, there is the average electric field $E_{\mathrm{avg}}$ between the needle tip and the grounded electrode (as used in e.g.~\cite{seeger2017, seeger2018}):
\begin{equation}
  \label{eq:average-field}
  E_\mathrm{avg} = |V| / d_\mathrm{gap}\,,
\end{equation}
where $d_\mathrm{gap} = 8$\,mm.
In our simulations we therefore have $E_{\mathrm{avg}} = 1.25 E_\mathrm{bg}$.

Note that the computational domain is relatively narrow.
Since we use Neumann boundary conditions for the electric potential, the discharge will develop as if identical discharges were developing around it, by mirroring the computational domain in the $x$ and $y$ directions.
This tends to somewhat artificially confine the discharge in the $x$ and $y$ directions.
However, we opted for such a domain since it reduces the rather high computational costs of the simulations.

\subsection{2D computational domain}\label{sec:comput-domain-2d}

As will be discussed in section~\ref{sec:pos-3d-results}, we are only able to simulate the early inception stage of positive discharges in 3D.
To qualitatively illustrate how such positive discharges could develop, we perform 2D Cartesian simulations using the computational domain shown in figure~\ref{fig:computational_domain}(b).
This domain measures (10\,mm)$^2$ and has the same electrode geometry and boundary conditions as the 3D domain discussed above.

\subsection{Included free electron sources}\label{sec:free-electron-sources}

In the simulations, we consider two sources of (initial) free electrons: a plasma seed near the electrode tip or stochastic background ionization, which will be explained below.
With a plasma seed, we are able to start negative discharges that continue to propagate, but not positive ones.
Positive discharges require a source of free electrons ahead of them for their propagation, but there is considerable uncertainty in the mechanisms that could provide such free electrons in C$_4$F$_7$N-CO$_2$ mixtures.

Two potential free electron sources are background ionization and photoionization.
Due to the fast attachment of electrons to C$_4$F$_7$N, background ionization would be present in the form of positive and negative ions.
Although there is evidence of electron detachment in high electric fields and low pressures in pure C$_4$F$_7$N~\cite{hosl2019}, it is not yet known which ions would form in C$_4$F$_7$N-CO$_2$ mixtures at atmospheric pressure and how easily electrons would detach from these ions.
Even less is known about photoionization in these mixtures.
However, since photoionization in CO$_2$ is much weaker than in air, due to the lower production of ionizing photons and their significantly smaller absorption distance~\cite{pancheshnyi2014}, we expect photoionization to be weak in C$_4$F$_7$N-CO$_2$ mixtures as well.

Due to the above uncertainties we decided to include a simple stochastic background ionization process in some of our simulations, to qualitatively illustrate how discharges would develop with a continuous free electron source.
Electron-ion pairs are produced in the whole computational domain at a rate $k_\mathrm{0}$ between $10^{18} \, \textrm{m}^{-3}\textrm{s}^{-1}$ and $10^{19} \, \textrm{m}^{-3}\textrm{s}^{-1}$.
Note that these amounts correspond to $1 \, \textrm{mm}^{-3}\textrm{ns}^{-1}$ and $10 \, \textrm{mm}^{-3}\textrm{ns}^{-1}$, respectively.
This production rate is high compared to laboratory experiments with a radioactive admixture~\cite{nijdam2011}, but much lower than the production rate of photoelectrons in air-like mixtures (for reference, in atmospheric air the photoionization rate is typically between $10^{22} \, \textrm{m}^{-3}\textrm{s}^{-1}$ and $10^{24} \, \textrm{m}^{-3}\textrm{s}^{-1}$ one millimeter away from a streamer discharge).
% We remark that this process contains physically realistic fluctuations as the number of electrons per cell is Poisson-distributed.

In the rest of our simulations (without stochastic background ionization), we only include an initial neutral seed near the tip of the electrode.
Electron-ion pairs are then generated according to a Gaussian distribution as
\begin{equation}
\label{eq:initial-seed}
    n_\mathrm{e}(\boldsymbol{\mathrm r}) = n_\mathrm{i}^{+}(\boldsymbol{\mathrm r}) = N_\mathrm{0} \, \mathrm{exp} \, \left[-\frac{(\boldsymbol{\mathrm r} - \boldsymbol{\mathrm r_0})^2}{\sigma^2}\right]\,,
\end{equation}
where $N_\mathrm{0} = 2 \times 10^{11} \, \textrm{m}^{-3}$ (unless specified otherwise), $\sigma = 0.2$\,mm, and $\boldsymbol{\mathrm r_0}$ is the position of the electrode tip located at $z = 8$\,mm.
The expected number of electrons in such a seed is $\pi^{3/2} \sigma^3 N_\mathrm{0}$, which is about 9 with the above values.

\section{Negative streamers in C$_4$F$_7$N-CO$_2$ mixtures}\label{sec:neg-streamers}

% In this section, we study negative streamers in C$_4$F$_7$N-CO$_2$ mixtures using 3D simulations.
% The simulations in section~\ref{sec:neg-initial-seed} were performed using an initial neutral seed, whereas stochastic background ionization was included in section~\ref{sec:neg-background-ionization}.
% We always performed multiple runs, but since results were generally rather similar between these runs, we only present an illustrative example for each simulation case.

\subsection{Streamers with an initial neutral seed}\label{sec:neg-initial-seed}

We first present 3D simulation results of negative streamers that start from an initial neutral seed placed near the electrode tip.
% The sensitivity of the simulations to this seed is discussed in~\ref{sec:neg-effect-of-N0}.
% In the simulations we vary the C$_4$F$_7$N concentration and the background electric field $E_\mathrm{bg}$ from their default values of 1\% and 36\,kV/cm, respectively.

\subsubsection{Effect of the background electric field}\label{sec:neg-effect-of-Ebg}
\begin{figure*}[t]
    \centering
    \includegraphics[width=1\textwidth]{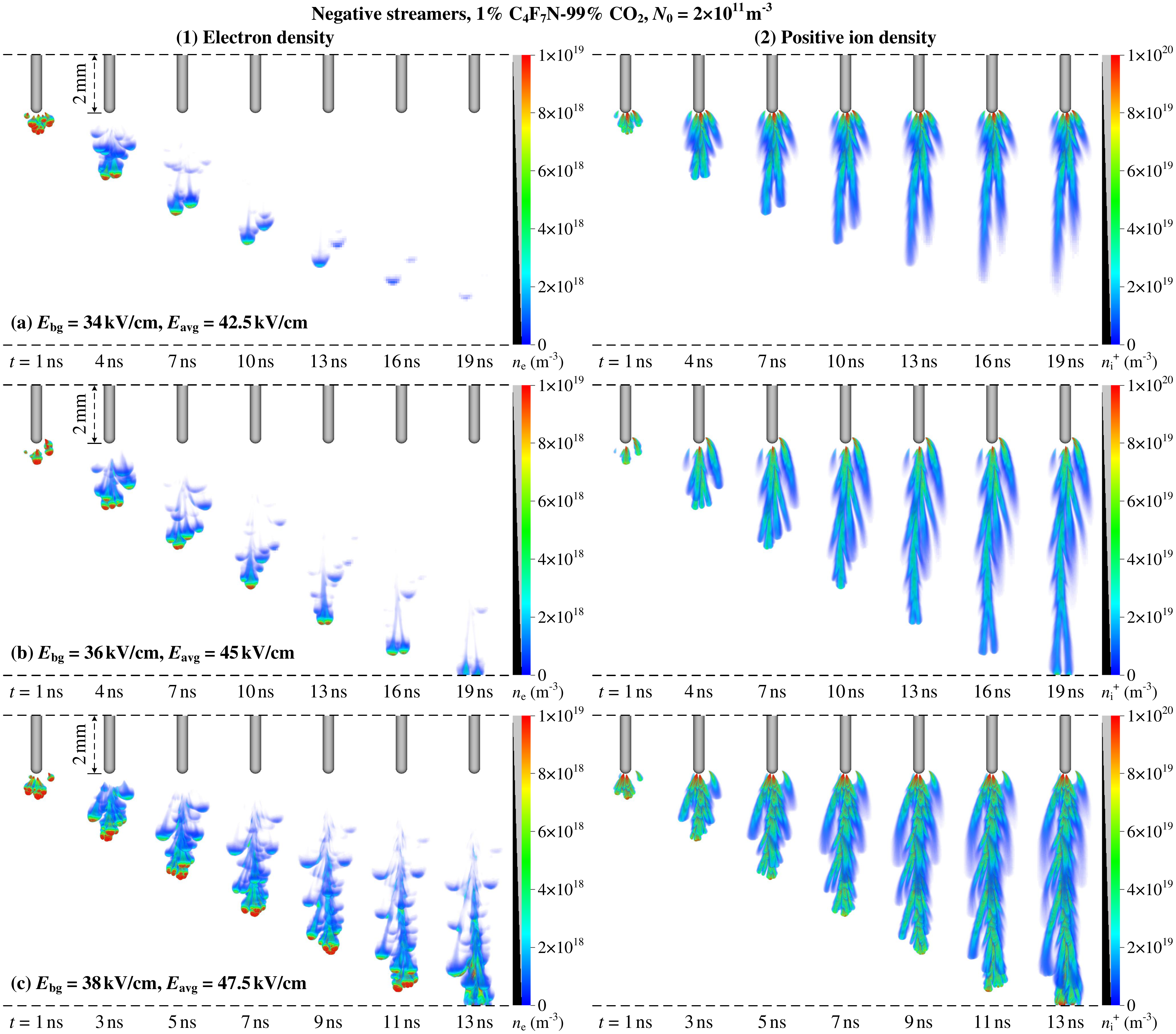}
    \caption{Effect of the background electric field $E_\mathrm{bg}$ on the propagation of negative streamers in a 1\% C$_4$F$_7$N-99\% CO$_2$ mixture with an initial neutral seed using 3D simulations.
    The upper and lower plate electrodes are here and afterward respectively indicated by a dashed line.
    Shown is the time evolution of (1) the electron density $n_\mathrm{e}$ and (2) the positive ion density $n_\mathrm{i}^{+}$ through Visit's~\cite{HPV:VisIt} 3D volume rendering; the opacity is indicated in the legend.
    The same visualization is also applied to subsequent 3D simulations.
    % from the side view, with a linear scale ranging from 0 to $1\times10^{19}\,\mathrm{m}^{-3}$ for $n_\mathrm{e}$ and a linear scale ranging from 0 to $4\times10^{20}\,\mathrm{m}^{-3}$ for $n_\mathrm{i}^{+}$
    A short conductive channel is observed behind the streamer head due to the fast attachment of electrons to C$_4$F$_7$N, so that the background field required to cross the gap is approximately the critical field $E_\mathrm{k} \approx 40$\,kV/cm.
    We remark that the maximal electron density is about $10^{20}\,\mathrm{m}^{-3}$, which is well above the limit of the color scale.}
    \label{fig:neg_different_E}
    % $E_\mathrm{max}$ at the electrode tip is about $4 \times 10^7$\,V/m.
    % $E_\mathrm{max}$ at the streamer head is about $1\sim1.5 \times 10^7$\,V/m.
    % The maximal electron density $n_\mathrm{e}$ is about $2\sim5 \times 10^{20}\,\mathrm{m}^{-3}$.
\end{figure*}

% $\quad$
Figure~\ref{fig:neg_different_E} illustrates the effect of the background electric field $E_\mathrm{bg}$ on negative streamers in a 1\% C$_4$F$_7$N-99\% CO$_2$ mixture.
First, in figure~\ref{fig:neg_different_E}(b) we look into the case of a background field of $E_\mathrm{bg} = 36$\,kV/cm, which corresponds to $E_\mathrm{avg} = 45$\,kV/cm.
For comparison, the critical field $E_\mathrm{k}$ is about 40\,kV/cm.
At $t = 1$\,ns, electrons provided by the initial neutral seed initiate a streamer near the electrode due to the locally high electric field.
The negative discharge grows at an almost constant velocity of about $0.5 \times 10^6$\,m/s until it crosses the gap at $t = 19$\,ns.
Multiple branches form during its propagation.

% \begin{table*}
% \renewcommand{\arraystretch}{1.2}
% \centering
% \captionsetup{width=1.0\textwidth}
% \caption{The axial electric field $E_\mathrm{ax}(z)$ at which $z = 0$\,mm (the lower plate) for three simulation cases corresponding to figure~\ref{fig:neg_different_E}, in units of kV/cm.}
% \label{tab:Eax(0)-three-cases}
% \begin{tabular*}{1.0\textwidth}{l@{\extracolsep{\fill}}ccccccc}
%   \br
%   Simulation case & $t = 0$\,ns & $t = 1$\,ns & $t = 4$\,ns & $t = 7$\,ns & $t = 10$\,ns & $t = 13$\,ns & $t = 16$\,ns \\
%   \mr
%   $E_\mathrm{bg} = 34$\,kV/cm, $E_\mathrm{avg} = 42.5$\,kV/cm & 35.4 & 35.7 & 36.3 & 36.5 & 36.7 & 36.7 & 36.8 \\
%   $E_\mathrm{bg} = 36$\,kV/cm, $E_\mathrm{avg} = 45.0$\,kV/cm & 37.5 & 37.7 & 38.2 & 38.5 & 38.8 & 39.4 & 42.4 \\
%   $E_\mathrm{bg} = 38$\,kV/cm, $E_\mathrm{avg} = 47.5$\,kV/cm & 39.6 & 39.9 & 40.8 & 41.7 & 46.1 & -- & -- \\
%   \br
% \end{tabular*}
% \end{table*}

\begin{figure}
    \centering
    \includegraphics[width=0.49\textwidth]{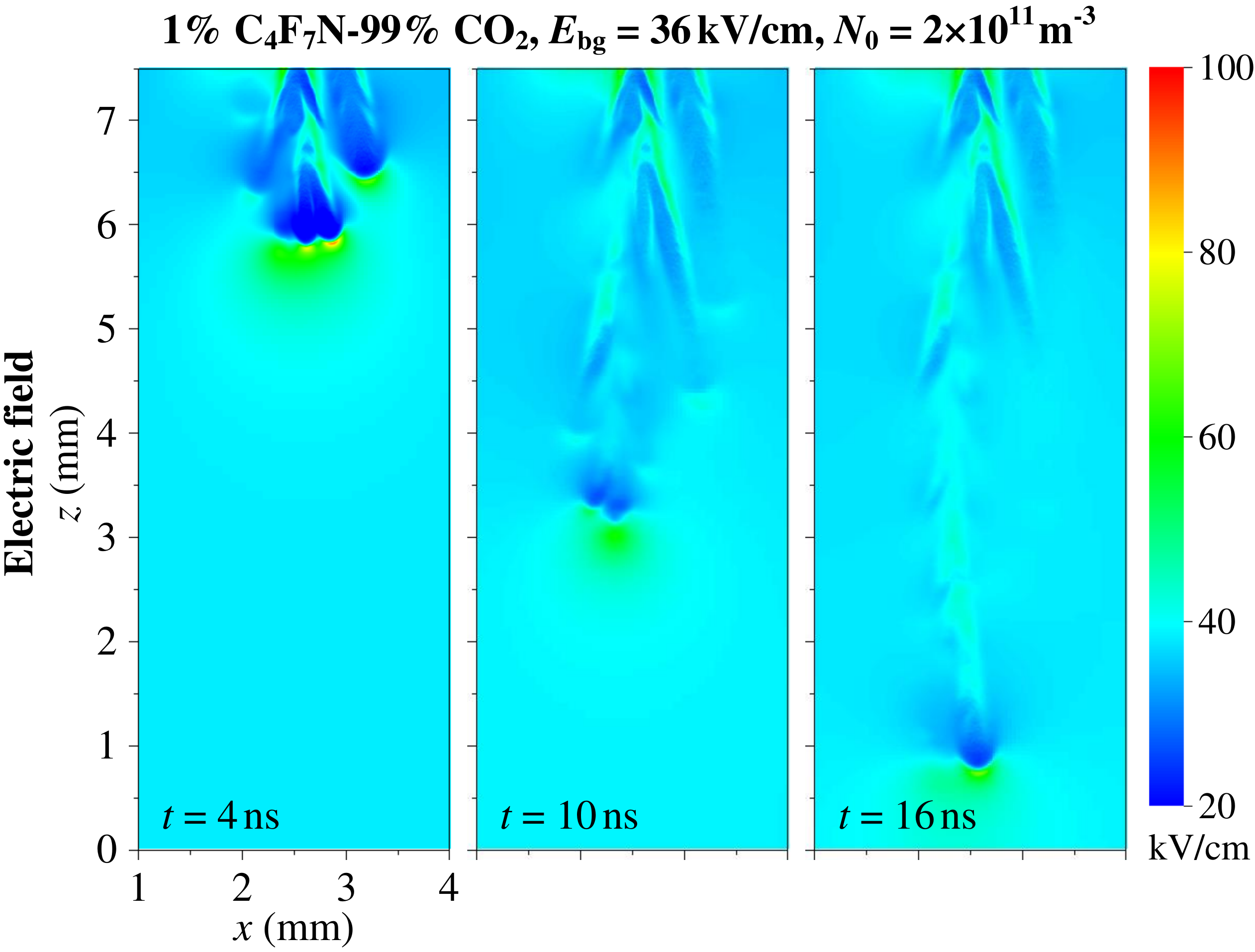}
    \caption{Cross sections of the electric field $E$ for the negative streamer shown in figure~\ref{fig:neg_different_E}(b) at $E_\mathrm{bg}=36$\,kV/cm.
    The electric field behind the streamer head approximately relaxes back to the background electric field due to rapid electron attachment.
    Note that the color scale ranges from 20\,kV/cm to 100\,kV/cm.}
    \label{fig:neg_3d_cross_sections}
\end{figure}

Electrons are primarily observed around the streamer head, where the maximal electron density is about $10^{20}\,\mathrm{m}^{-3}$, as those inside the streamer channel rapidly attach to C$_4$F$_7$N.
Only a short part of the channel behind the streamer head therefore has a significant electron conductivity, with the electric field in the channel approximately relaxing back to the background electric field, as shown in figure~\ref{fig:neg_3d_cross_sections}.
To better show the discharge trajectory and morphology, the positive ion density $n_\mathrm{i}^{+}$ is also included in figure~\ref{fig:neg_different_E} as well as subsequent figures.
Note that the color scale in these visualizations can affect apparent width of the channels, as illustrated in~\ref{sec:neg-effect-of-color-scale}.

In figure~\ref{fig:neg_different_E}(a), the background electric field is decreased to $E_\mathrm{bg} = 34$\,kV/cm, corresponding to $E_\mathrm{avg} = 42.5$\,kV/cm.
The negative discharge now decelerates and fades out before reaching the lower plate, losing its field enhancement, similar to fading negative streamers in air~\cite{guo2022d}.
This is initially surprising, since the average electric field $E_\mathrm{avg} = 42.5$\,kV/cm exceeds $E_\mathrm{k}$.
However, this phenomenon can be explained by the fast decay of the electron conductivity in the channels, which leads to streamers with weak field enhancement and poorly conducting channels, as illustrated in figure~\ref{fig:neg_3d_cross_sections}.
Streamers in strongly attaching gases thus modify the background electric field to a much smaller extent than streamers in gases such as air, which means that their propagation requires a background electric field close to the critical field $E_\mathrm{k}$.
When $E_\mathrm{avg} \approx E_\mathrm{k}$, there is a region where the background field exceeds $E_\mathrm{k}$ near the rod electrode but also a region below $E_\mathrm{k}$ farther away from it, and streamers can stop propagating in the latter region, as shown in figures~\ref{fig:neg_different_E}(a).

% Negative streamers can only keep propagating if the background field $E_\mathrm{bg}$ is close to the critical field $E_\mathrm{k} \approx 40$\,kV/cm, as shown in figures~\ref{fig:neg_different_E}(b) and~\ref{fig:neg_different_E}(c).

% \st{In other words, the streamer stability field $E_\mathrm{st}$~\cite{allen1991} (i.e., the background electric field required for a streamer to cross a gap) is approximately the critical electric field $E_\mathrm{k}$, as has also been observed in the electronegative gas SF$_6$~\cite{bujotzek2015}. 
% The reason for this is that fast electron attachment shortens the conductive channel behind the streamer head, which strongly reduces the electric field enhancement~\cite{guo2022d}.}

% Therefore, with a higher $E_\mathrm{bg}$, the negative streamer propagates faster and is more likely to cross the gap, forming a longer conductive channel and more complex structures.

\subsubsection{Effect of the initial neutral seed}\label{sec:neg-effect-of-N0}

The streamers cross the gap in a background field $E_\mathrm{bg} \approx 36$\,kV/cm. 
This value is close to the critical field $E_\mathrm{k}$, as has also been observed in the electronegative gas SF$_6$~\cite{bujotzek2015}.  
The reason is that fast electron attachment shortens the conductive channel behind the streamer head, which strongly reduces the electric field enhancement at the head. (For comparison, the effect of strong attachment on axisymmetric streamers in air was studied in~\cite{francisco2021a}.)

\begin{figure*}[t]
    \centering
    \includegraphics[width=1\textwidth]{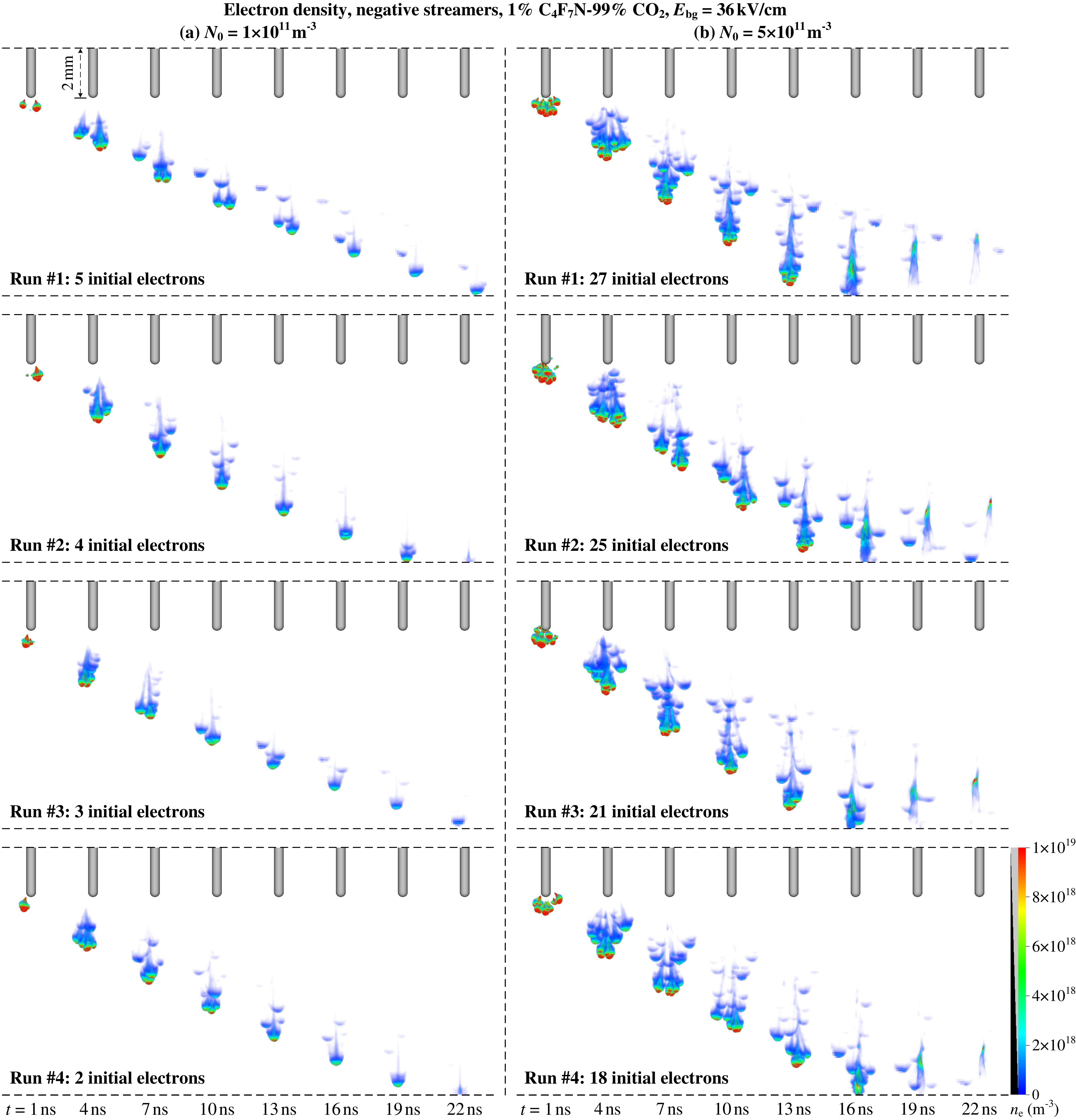}
    \caption{Effect of the maximal electron density $N_\mathrm{0}$ of the initial seed on the propagation of negative streamers. 
    All parameters are the same as in  figure~\ref{fig:neg_different_E}(b), except for $N_\mathrm{0}$, which is changed from $2 \times 10^{11} \, \textrm{m}^{-3}$ to $1 \times 10^{11} \, \textrm{m}^{-3}$ and $5 \times 10^{11} \, \textrm{m}^{-3}$, respectively.
    For both cases, four simulation runs with different numbers of initial electrons are shown.
    An increase of the density $N_\mathrm{0}$ leads to the formation of a wider and faster discharge containing more streamer channels, making it easier for the discharge to cross the gap.
    This resembles the observation that in air streamers with larger radii propagate steadily in lower background fields~\cite{li2022a,guo2022d}.}
    \label{fig:neg_different_sd}
    % $E_\mathrm{max}$ at the electrode tip is about $4 \times 10^7$\,V/m.
    % $E_\mathrm{max}$ at the streamer head is about $1\sim1.2 \times 10^7$\,V/m.
    % The maximal electron density $n_\mathrm{e}$ is about $2\sim4 \times 10^{20}\,\mathrm{m}^{-3}$.
\end{figure*}

From these results, one might conclude that the stability field $E_\mathrm{st}$ is 36\,kV/cm, if the stability field is defined as the background electric field required for a streamer to cross a gap~\cite{allen1991}. 
% {\it IS THIS THE GOOD DEFINITION? COMPARE WITH DISCUSSION AND REFERENCES IN~\cite{guo2022d}.} 
However, we have shown recently in~\cite{li2022a, guo2022d} that the electric field where streamers in air propagate steadily, i.e.\ with constant shape and velocity, depends on their radius. 
So there is no unique stability field, but streamers with larger radii propagate steadily in lower background electric fields. 
A similar effect can be seen in figure~\ref{fig:neg_different_sd}.
Here the same conditions are used as in figure~\ref{fig:neg_different_E}(b), except that the maximal electron density $N_\mathrm{0}$ of the initial seed is changed from $2 \times 10^{11} \, \textrm{m}^{-3}$ to $1 \times 10^{11} \, \textrm{m}^{-3}$ and $5 \times 10^{11} \, \textrm{m}^{-3}$, respectively. 
% In the first case (figure~\ref{fig:neg_different_sd}(a)), the weaker initial ionization launches fewer streamers that do not cross the gap, while the larger seed creates a wider discharge containing more streamers that collectively cross the gap, in the same background field of 36\,kV/cm.
Although most discharges are able to cross the gap, we find that the larger seed creates a wider discharge containing more streamer channels, making it easier for the discharge to cross the gap in the same background field of $E_\mathrm{bg} = 36$\,kV/cm.
% The average streamer velocity of $0.55 \times 10^6$\,m/s is also higher (compared to $0.4 \times 10^6$\,m/s for figure~\ref{fig:neg_different_sd}(a)).
The average streamer velocity is also about 40\% higher.
This is due to a collective effect where neighboring branches mutually increase the electric field enhancement at the streamer head.

% \begin{figure*}
%     \centering
%     \includegraphics[width=1\textwidth]{Figures/neg_0.01eco_3.6e6_different_sd.pdf}
%     \caption{Effect of the maximal electron density $N_\mathrm{0}$ of the initial seed on the propagation of negative streamers. 
%     All parameters are the same as in  figure~\ref{fig:neg_different_E}(b), except for $N_\mathrm{0}$, which is changed from $2 \times 10^{11} \, \textrm{m}^{-3}$ to $1 \times 10^{11} \, \textrm{m}^{-3}$ and $5 \times 10^{11} \, \textrm{m}^{-3}$, respectively.
%     % The maximal electron density is about $10^{20}\,\mathrm{m}^{-3}$. 
%     An increase of this density $N_\mathrm{0}$ leads to the formation of a wider discharge containing more streamer channels, making it easier for the discharge to cross the gap.
%     This resembles the observation that in air streamers with larger radii propagate steadily in lower background fields~\cite{li2022a,guo2022d}.}
%     \label{fig:neg_different_sd}
%     % $E_\mathrm{max}$ at the electrode tip is about $4 \times 10^7$\,V/m.
%     % $E_\mathrm{max}$ at the streamer head is about $1\sim1.2 \times 10^7$\,V/m.
%     % The maximal electron density $n_\mathrm{e}$ is about $2\sim4 \times 10^{20}\,\mathrm{m}^{-3}$.
% \end{figure*}

\subsubsection{Effect of the C$_4$F$_7$N concentration}\label{sec:neg-effect-of-C4F7N-concentration}

\begin{figure*}
    \centering
    \includegraphics[width=1\textwidth]{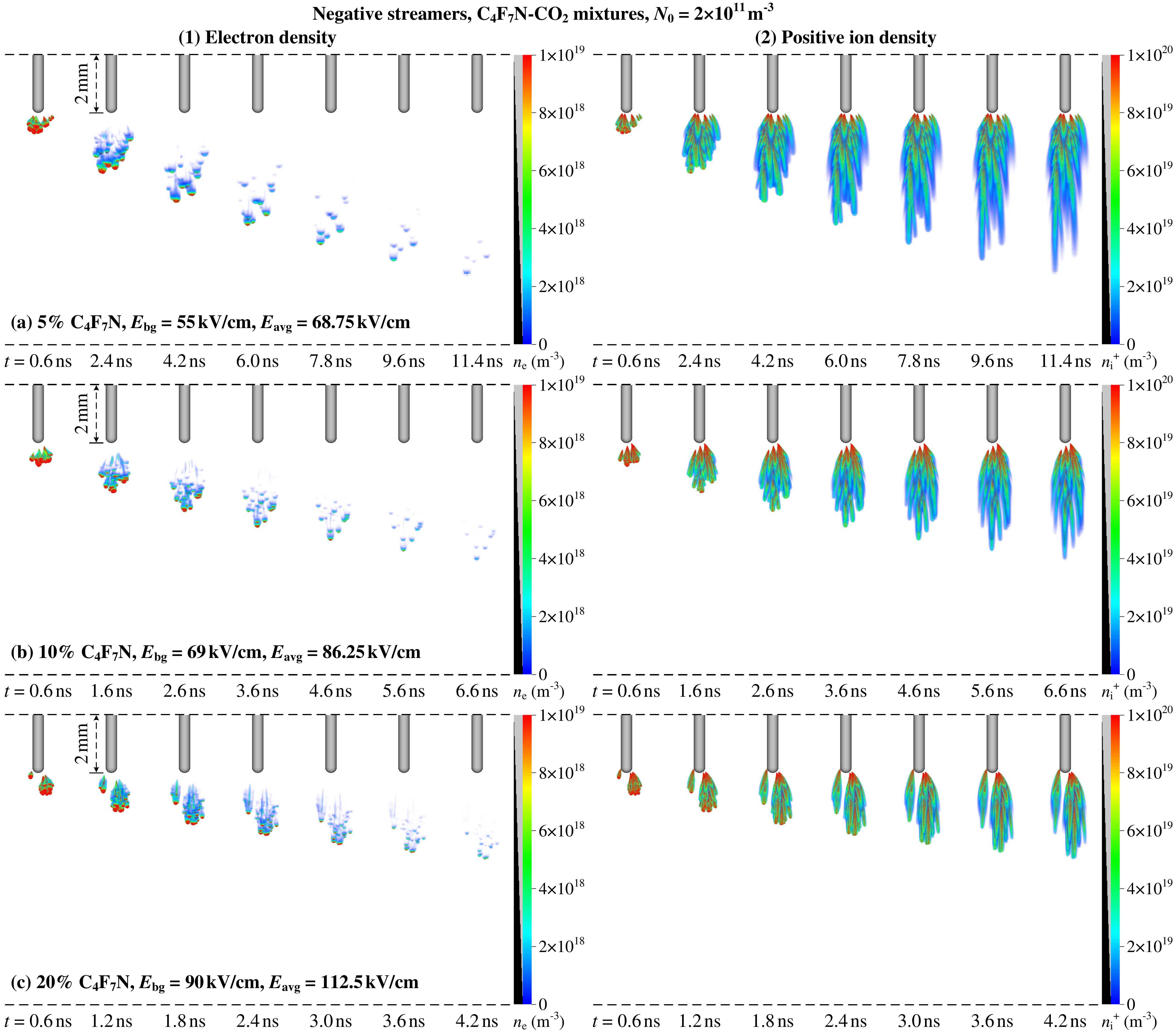}
    \caption{Effect of the C$_4$F$_7$N concentration on the propagation of negative streamers at $E_\mathrm{bg} = 0.9 \, E_\mathrm{k}$ with an initial neutral seed, see table~\ref{tab:critical-field}.
    The maximal electron density is about $10^{21}\,\mathrm{m}^{-3}$.
    As the C$_4$F$_7$N concentration increases, the streamer branches into more channels, which are thinner, slower, and stop earlier.}    \label{fig:neg_different_C4F7N_concentration}
    % $E_\mathrm{max}$ at the electrode tip is about $6\sim10 \times 10^7$\,V/m.
    % $E_\mathrm{max}$ at the streamer head is about $1.5\sim2.5 \times 10^7$\,V/m.
    % The maximal electron density $n_\mathrm{e}$ is about $1\sim3 \times 10^{21}\,\mathrm{m}^{-3}$.
\end{figure*}

The effect of the C$_4$F$_7$N concentration is illustrated in figure~\ref{fig:neg_different_C4F7N_concentration}, where the C$_4$F$_7$N fraction is increased to 5\%, 10\% and 20\%.
These mixtures are of interest in practical applications as their dielectric strength is comparable to SF$_6$.
Given the sensitivity of negative streamer propagation to the background electric field $E_\mathrm{bg}$, here we set $E_\mathrm{bg} = 0.9 \, E_\mathrm{k}$ for all cases, see table~\ref{tab:critical-field}.

All negative streamers decelerate and fade out before crossing the gap, in contrast to the case with 1\% C$_4$F$_7$N shown in figure~\ref{fig:neg_different_E}(b), which is also at $E_\mathrm{bg} = 0.9 \, E_\mathrm{k}$.
Furthermore, the streamer channels are thinner, slower and they branch more when the C$_4$F$_7$N fraction is increased, and they stop earlier.
This indicates that for increased C$_4$F$_7$N concentrations the background electric field $E_\mathrm{bg}$ required to cross the gap is above $0.9 \, E_\mathrm{k}$, due to a higher electron attachment rate that further decreases the streamer's electric field enhancement.

% Finally, we remark that the maximal electron density increases with the C$_4$F$_7$N fraction (up to $10^{21}\,\mathrm{m}^{-3}$).

\subsection{Streamers with stochastic background ionization}\label{sec:neg-background-ionization}

\begin{figure*}
    \centering
    \includegraphics[width=1\textwidth]{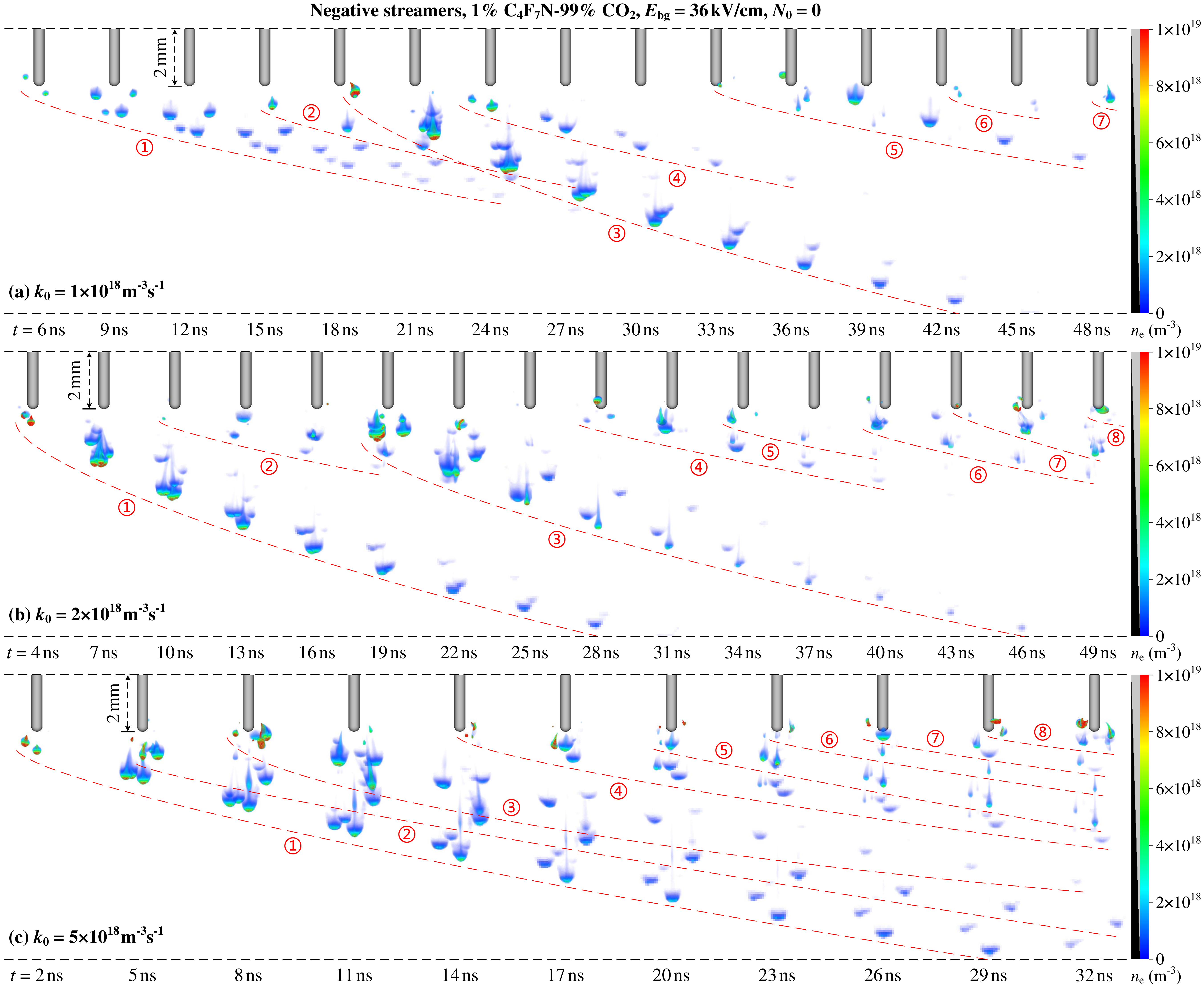}
    \caption{Effect of the background ionization rate $k_\mathrm{0}$ on the propagation of negative streamers without an initial neutral seed.
    The maximal electron density is about $10^{20}\,\mathrm{m}^{-3}$.
    With stochastic background ionization, a remarkable phenomenon occurs where new negative streamers are generated behind the previous ones, due to rapid electron attachment.
    This process repeats itself leading to the formation of a chain of negative streamers, each represented by a red dashed curve and sequentially numbered to indicate their order of emergence.}
    \label{fig:neg_different_bir}
    % $E_\mathrm{max}$ at the electrode tip is about $4 \times 10^7$\,V/m.
    % $E_\mathrm{max}$ at the streamer head is about $1 \times 10^7$\,V/m.
    % The maximal electron density $n_\mathrm{e}$ is about $2\sim6 \times 10^{20}\,\mathrm{m}^{-3}$.
\end{figure*}

We now perform 3D simulations of negative streamers in which the initial neutral seed is replaced by stochastic background ionization that continuously produces electron-ion pairs within the computational domain, as discussed in section~\ref{sec:free-electron-sources}.
Although such background ionization is not required for negative streamer propagation, we do observe some interesting effects in figure~\ref{fig:neg_different_bir}, in which the background ionization rate $k_\mathrm{0}$ is varied.
With a higher $k_\mathrm{0}$ more electron-ion pairs are produced per unit of time.
As expected, this results in earlier initiation and the formation of a greater number of streamers.

A remarkable phenomenon is that new negative streamers are generated behind the previous ones.
This happens because the electric field behind previous streamers quickly relaxes back to the background electric field, due to rapid electron attachment, which allows new streamers to form.
This process repeats itself, at least within the time scales considered, so that a chain of negative streamers emerges.
In some cases, later streamers are faster and they can approach or overtake previous ones, as shown in figure~\ref{fig:neg_different_bir}(a) for the streamers numbered (2) and (3).
% \ue{\it SHOULD WE STOP THE SECTION HERE AND NOT SPECULATE ABOUT THE END OF THE CHAIN? I AM NOT SURE ABOUT THE STATEMENTS BELOW. E.G., GLOBALLY THERE IS CHARGE NEUTRALITY: THERE IS AN EQUAL NUMBER OF POSITIVE AND NEGATIVE IONS, IF ELECTRODE CURRENTS ARE NEGLECTED. BUT ELECTRONS FLOW FROM THE UPPER ELECTRODE INTO THE GAS, IONIZE IT AND LEAVE MORE POSITIVE THAN NEGATIVE IONS BEHIND. THESE POSITIVE CHARGES DRIFT UPWARDS $\ldots$ EVENTUALLY YOU WILL GET SOMETHING LIKE TRICHEL PULSES, I.E, A CHAIN OF PULSES ON A MUCH LONGER TIME SCALE?}
Although there is only a low electron conductivity behind the repeated streamers, the ion density and hence the ion conductivity increase over time, if the recombination of positive and negative ions is not too fast. 
If the ion conductivity keeps increasing, we expect that the ions eventually will screen the electric field and inhibit the formation of new streamers.
% \st{We expect that this will eventually cause the
% chain to stop, due to electric field screening by these ions, which will inhibit the formation of new streamers.}

\section{Positive streamers in C$_4$F$_7$N-CO$_2$ mixtures}\label{sec:pos-streamers}

In this section, we perform both 3D and 2D simulations of positive streamers in C$_4$F$_7$N-CO$_2$ mixtures.
The simulations include stochastic background ionization, which serves as the source of free electrons for positive streamer propagation, as discussed in section~\ref{sec:free-electron-sources}.

\subsection{3D simulation results}\label{sec:pos-3d-results}

\begin{figure*}
    \centering
    \includegraphics[width=1\textwidth]{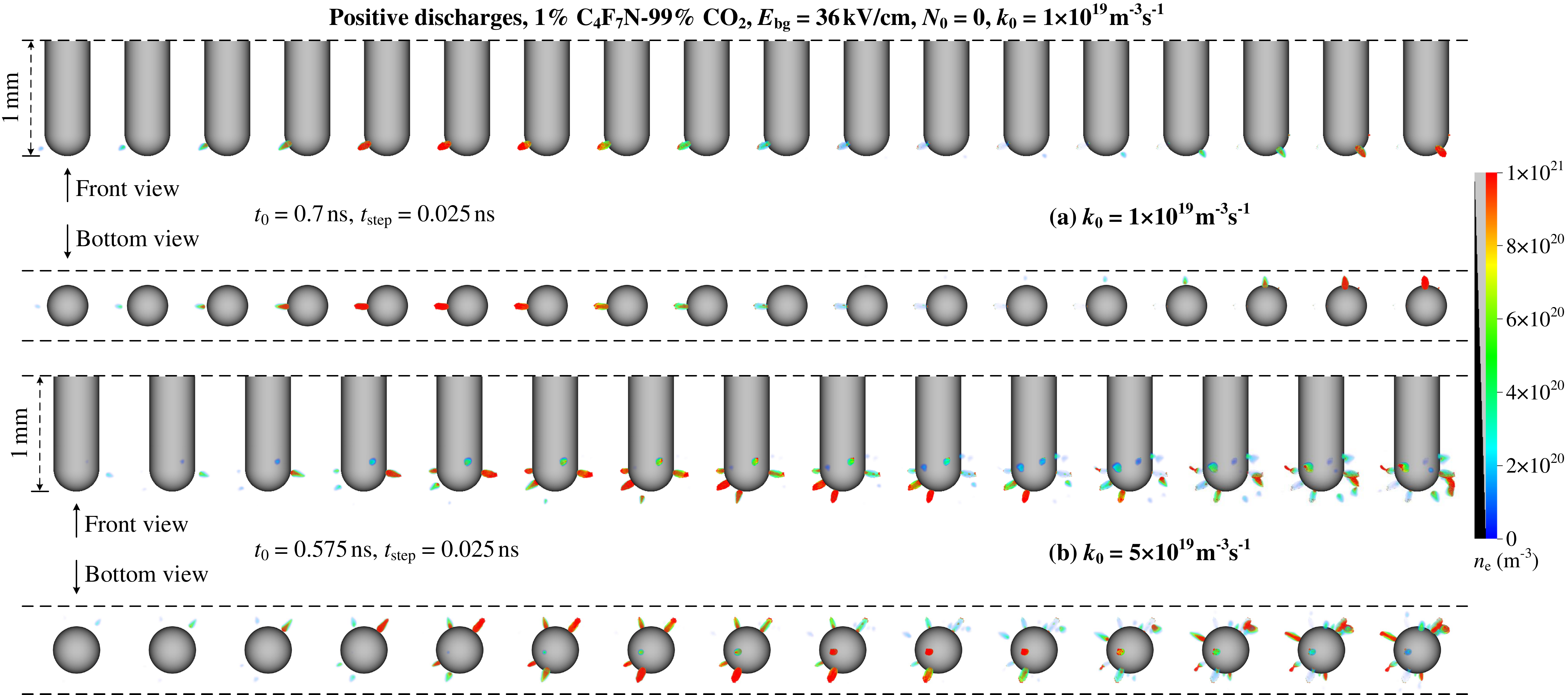}
    \caption{Two examples of positive discharges in 3D simulations using stochastic background ionization, without an initial neutral seed.
    Both the side and bottom views are presented, with a zoomed-in perspective centered at the electrode tip.
    Here we only show the time evolution of the electron density $n_\mathrm{e}$ in time steps of 0.025\,ns through Visit's 3D volume rendering, with a linear scale ranging from 0 to $1\times10^{21}\,\mathrm{m}^{-3}$.
    The maximal electron density is about $10^{24}\,\mathrm{m}^{-3}$.
    In 3D, we are only able to simulate the early inception stage of positive discharges due to the presence of extremely high electric fields and electron densities.}
    \label{fig:pos_3d_different_bir}
    % $E_\mathrm{max}$ at the head of the discharge is about $3\sim5 \times 10^8$\,V/m.
    % The maximal electron density $n_\mathrm{e}$ is about $1\sim2 \times 10^{24}\,\mathrm{m}^{-3}$.
\end{figure*}

\begin{figure*}
    \centering
    \includegraphics[width=1\textwidth]{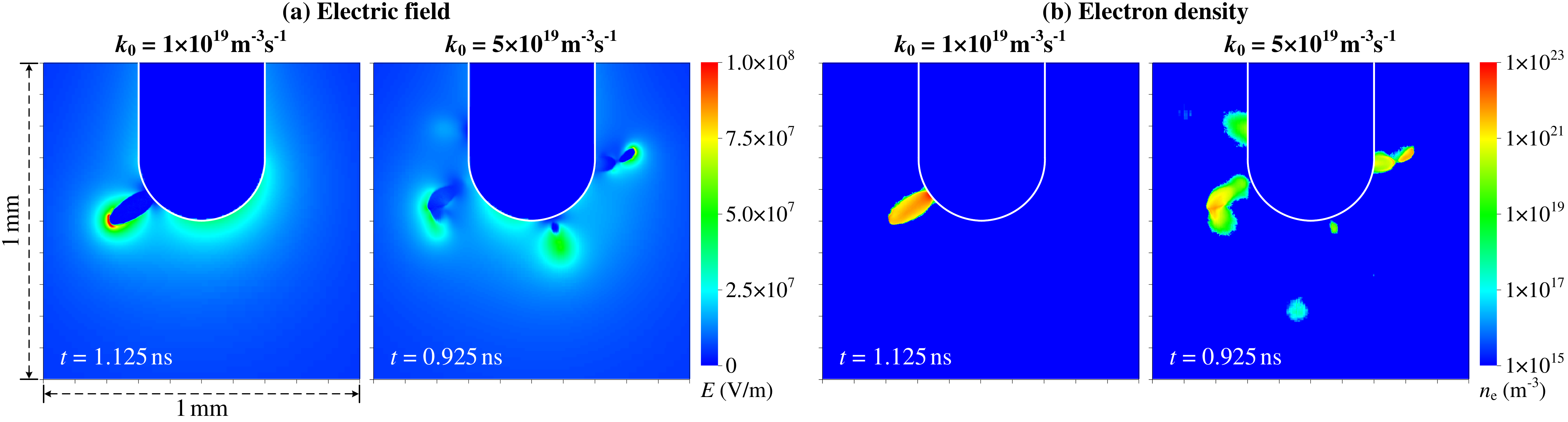}
    \caption{Cross sections of (a) the electric field $E$ and (b) the electron density $n_\mathrm{e}$ for the two positive discharges shown in figure~\ref{fig:pos_3d_different_bir} at the last frame, with a $(1\,\mathrm{mm})^{2}$ zoomed-in perspective centered at the electrode tip.
    Note that $n_\mathrm{e}$ is shown on a logarithmic scale.
    $E_\mathrm{max}$ exceeds $10^8$\,V/m, and the maximal electron density exceeds $10^{23}\,\mathrm{m}^{-3}$.}
    \label{fig:pos_3d_cross_sections}
\end{figure*}

In 3D, we can only simulate the early inception stage of positive discharges. 
Two examples for $k_\mathrm{0} = 1 \times 10^{19} \, \textrm{m}^{-3}\textrm{s}^{-1}$ and $k_\mathrm{0} = 5 \times 10^{19} \, \textrm{m}^{-3}\textrm{s}^{-1}$ are shown in figure~\ref{fig:pos_3d_different_bir}.
In these simulations, electron avalanches develop towards the electrode, which quickly transition into negative streamers.
Since these negative streamers are much thinner than the electrode itself, the maximal electric field at the streamer heads becomes extremely high, exceeding $10^8$\,V/m, as illustrated in figure~\ref{fig:pos_3d_cross_sections}.
This also results in very high electron densities, exceeding $10^{23}\,\mathrm{m}^{-3}$.

Computationally, it is very demanding to simulate discharges with such high fields and electron densities, due to the high grid resolution (below $1 \, \mu\textrm{m}$) and small time steps (below $10^{-14}$\,s) required.
Although we are only able to simulate the first stages of inception, our results suggest that positive discharges develop in a highly irregular way in C$_4$F$_7$N-CO$_2$ mixtures, with sharp features and very high fields at their tips.
We speculate that the growth of positive discharges will be the result of incoming negative streamers that connect to an existing discharge channel.
To qualitatively study such growth, we will perform 2D Cartesian simulations in section~\ref{sec:pos-2d-results}.

\subsection{2D Cartesian simulation results}\label{sec:pos-2d-results}

To qualitatively study the growth of positive discharges in C$_4$F$_7$N-CO$_2$ mixtures, we now perform 2D Cartesian simulations.
Computational costs are much lower in such a geometry, not only because there is one less dimension but also because field enhancement is significantly weaker, as there is no curvature in the third dimension.
Although there are quantitative differences between 2D and 3D in terms of streamer properties and branching, 2D simulations can help to qualitatively understand the growth of positive discharges.

The interpretation of particle weights and stochastic fluctuations is somewhat complicated in 2D.
We here use a minimum particle weight of $w_\mathrm{min} = 10^5 \, \mathrm{m}^{-1}$.
This can for example be interpreted as having an unresolved ``third'' dimension in the simulations of $1 \, \textrm{m}$ and a minimum particle weight of $10^5$, or equivalently as having and a third dimension of $10 \, \mu\textrm{m}$ and a minimum weight of one, see~\cite{li2023} for details.
We will express the background ionization rate as $k_\mathrm{0}/w_\mathrm{min}$, with units $\mathrm{m}^{-2}\mathrm{s}^{-1}$.

\begin{figure*}
    \centering
    \includegraphics[width=1.25\textwidth, angle=90]{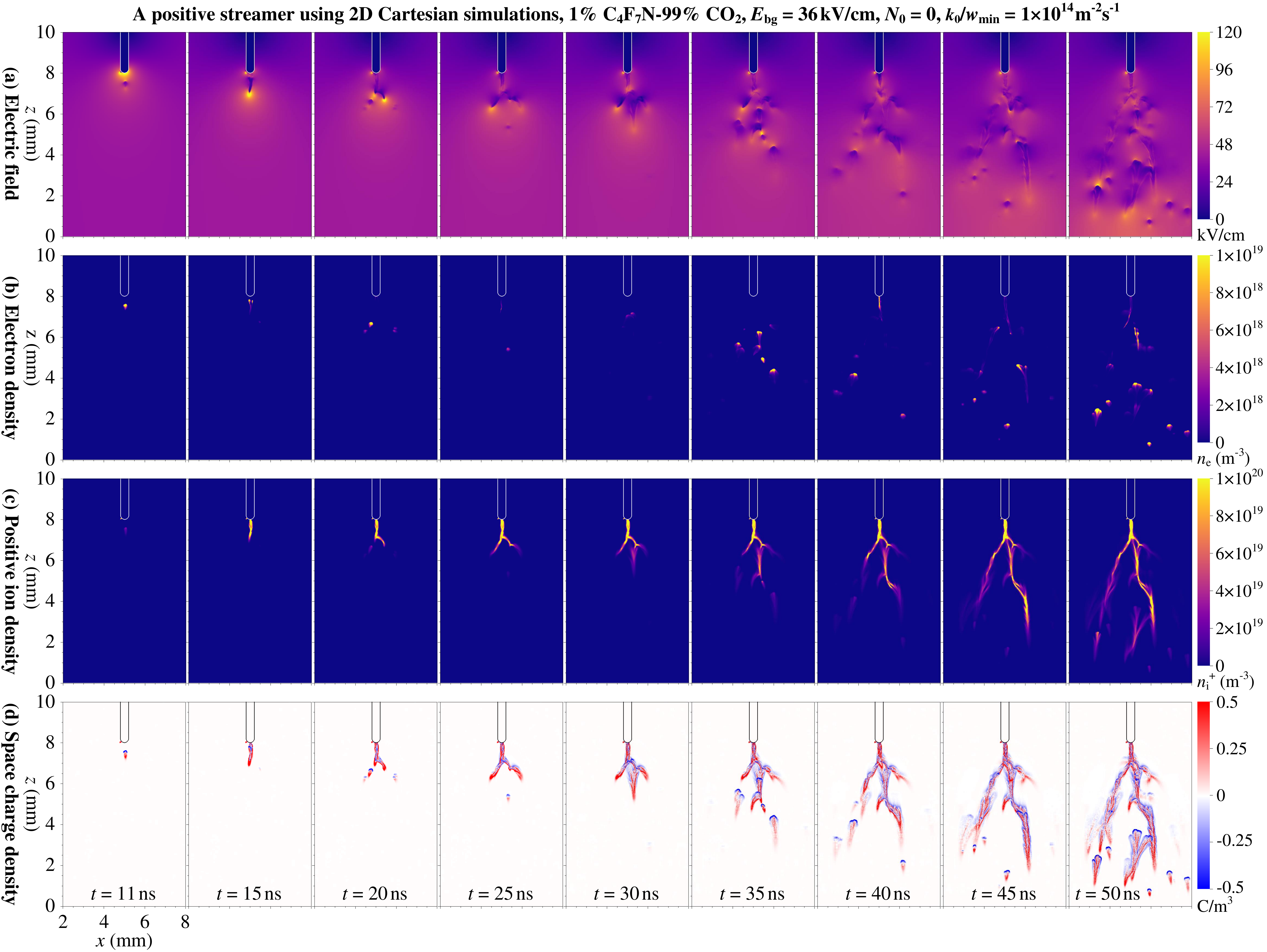}
    \caption{An example of a positive streamer using 2D Cartesian simulations with a background ionization rate $k_\mathrm{0}/w_\mathrm{min} = 1\times10^{14}\,\mathrm{m^{-2}s^{-1}}$, without an initial neutral seed.
    Shown is the time evolution of (a) the electric field $E$, (b) the electron density $n_\mathrm{e}$, (c) the positive ion density $n_\mathrm{i}^{+}$ and (d) the space charge density $\rho$.
    The growth of the positive discharge exhibits a highly irregular and branched structure, resulting from incoming negative streamers that connect to existing channels.}
    \label{fig:pos_2d_streamer_example}
    % $E_\mathrm{max}$ at the head of the discharge is about $1\sim2 \times 10^7$\,V/m.
    % The maximal electron density $n_\mathrm{e}$ is about $10^{19} \sim 10^{20}\,\mathrm{m}^{-3}$.
\end{figure*}

Figure~\ref{fig:pos_2d_streamer_example} shows the time evolution of a positive streamer in a 1\% C$_4$F$_7$N-99\% CO$_2$ mixture with $E_\mathrm{bg} = 36$\,kV/cm ($E_\mathrm{avg} = 45$\,kV/cm) and $k_\mathrm{0}/w_\mathrm{min} = 1\times10^{14}\,\mathrm{m^{-2}s^{-1}}$. 
Electron avalanches form due to stochastic background ionization, and they initially develop towards the electrode from various directions.
During their development, these avalanches transition into short negative streamers, which connect to each other, thereby extending the positive streamer channel.
This process repeats over time, leading to the irregular and branched downward propagation of the positive streamer.

\begin{figure*}
    \centering
    \includegraphics[width=1\textwidth]{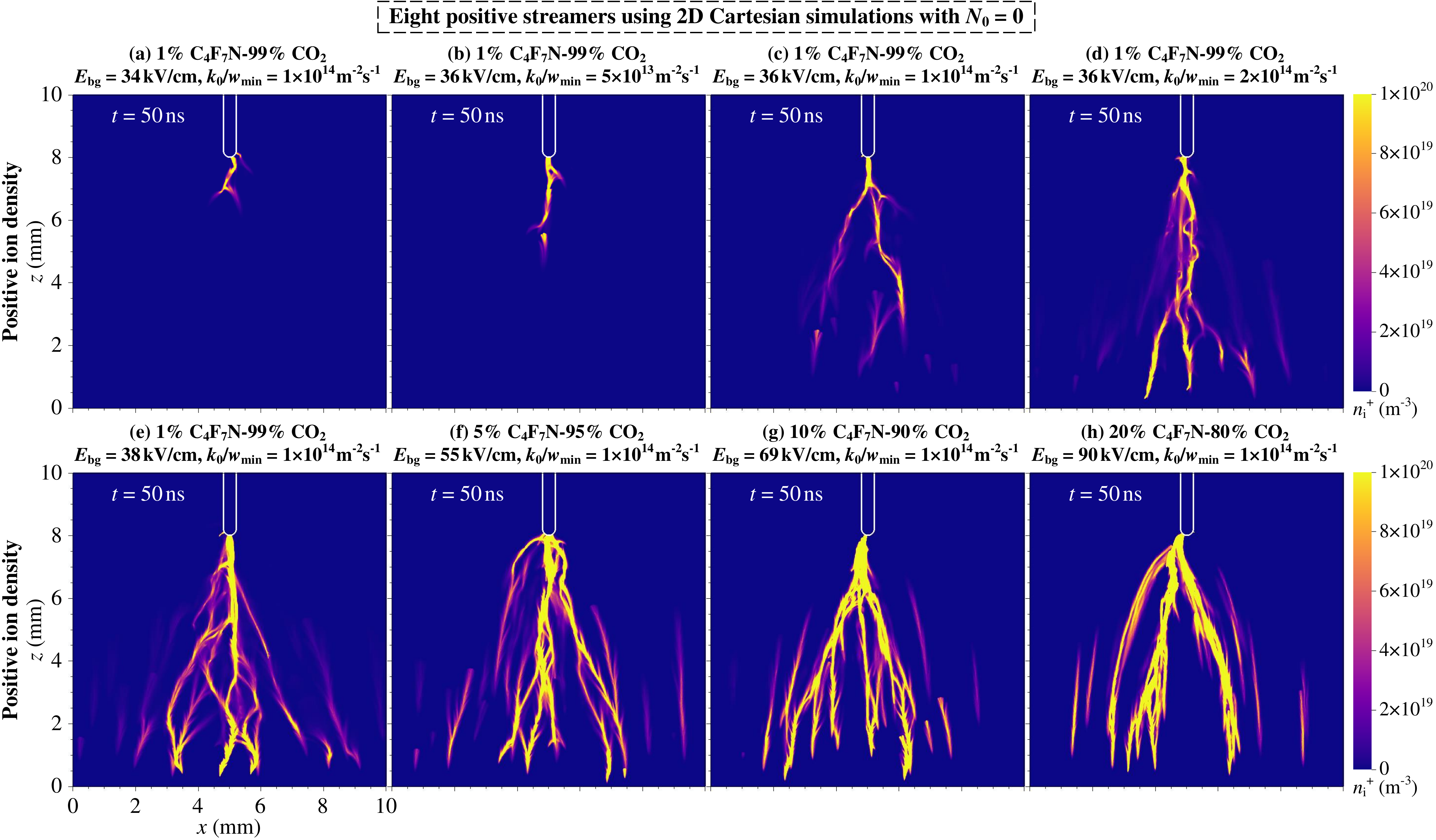}
    \caption{Positive ion density $n_\mathrm{i}^{+}$ for eight different positive streamers at $t=50$\,ns using 2D Cartesian simulations with stochastic background ionization, without an initial neutral seed.
    The background electric field $E_\mathrm{bg}$, background ionization rate $k_\mathrm{0}/w_\mathrm{min}$ and C$_4$F$_7$N concentration are varied.
    All the positive streamers develop in a highly irregular way, with the positive channels extending by incoming negative streamers.}
    \label{fig:pos_2d_eight_streamers}
    % $E_\mathrm{max}$ at the head of the discharge is about $1\sim3 \times 10^7$\,V/m.
    % The maximal electron density $n_\mathrm{e}$ is about $10^{19} \sim 10^{20}\,\mathrm{m}^{-3}$.
\end{figure*}

In figure~\ref{fig:pos_2d_eight_streamers} we present eight different positive streamers by varying the background electric field $E_\mathrm{bg}$, background ionization rate $k_\mathrm{0}/w_\mathrm{min}$ and C$_4$F$_7$N concentration.
Results are shown at $t=50$\,ns, and the streamers in panels (d)--(h) have crossed the gap.
In all cases, the positive channels extend by incoming negative streamers, as was illustrated by the temporal evolution in figure~\ref{fig:pos_2d_streamer_example}.
Note that the number of streamer channels increases with the background field and with the background ionization rate.

In 3D, we expect positive streamers to grow in a similar manner, with the channels extending by incoming negative streamers.
However, the electric fields at the streamer tips would be much higher, and so would be the electron densities.
Furthermore, one would generally expect more branching in 3D, although the number of branches would to some extent depend on the amount of background ionization.
Such highly stochastic growth has been experimentally observed in gases with weak or no photoionization~\cite{nijdam2010}.

\section{Comparison with past computational work}\label{sec:comparison-past-work}

Below, we briefly discuss related computational work.
To the best of our knowledge, all previous simulations of discharges in C$_4$F$_7$N mixtures were performed using 2D fluid models.
In~\cite{wang2021a}, streamers in C$_4$F$_7$N mixtures with different buffer gases were simulated at 300\,K and 1\,bar using an axisymmetric model~\cite{teunissen2017}.
Both ionization and attachment processes were considered, with their coefficients interpolated from~\cite{chachereau2018, long2019, long2020}.
Photoionization was not considered, but instead an initial homogeneous density of $10^{14} \, \textrm{m}^{-3}$ for electrons and positive ions was included.
The results showed that a higher background field was required for streamers to propagate in C$_4$F$_7$N-N$_2$ than in C$_4$F$_7$N-CO$_2$.

In~\cite{gao2022}, negative corona discharges in C$_4$F$_7$N-N$_2$ and C$_4$F$_7$N-CO$_2$ mixtures were simulated at 300\,K and 4\,bar using an axisymmetric model.
The ionization and attachment cross sections of C$_4$F$_7$N were obtained from~\cite{zhong2019a} and~\cite{chachereau2018}, respectively.
Electron-ion and ion-ion recombination processes were also considered. % assuming a constant reaction rate coefficient of $2.1\times10^{-13}\,\mathrm{m^3\,s^{-1}}$.
The negative corona discharges started from an initial Gaussian seed, and no other background ionization was included.
However, most of the negative discharges in the simulations remained in the inception stage and were far from propagation.

In~\cite{yan2023}, positive streamers in C$_4$F$_7$N were simulated in 2D with varying electrode shapes, using the same cross section data as~\cite{zhang2022b}.
An unspecified background ionization density was assumed as a replacement for photoionization.
Increasing the applied voltage led to higher maximum electric fields and velocities.
In another study~\cite{yan2023a}, the authors performed 2D simulations of surface discharges in a 9\% C$_4$F$_7$N-91\% CO$_2$ mixture at 300\,K and 1\,bar, again using an unspecified background ionization density.
They investigated the effect of the applied voltage and the dielectric constant on the discharge, and compared surface discharges in C$_4$F$_7$N-CO$_2$ with those in SF$_6$.

% In~\cite{zhang2022b}, particle swarm simulations were used to investigate the DC breakdown characteristics of C$_4$F$_7$N-CO$_2$ mixtures.
% Elastic, excitation, ionization and attachment collisions with C$_4$F$_7$N and CO$_2$ were included.
% For C$_4$F$_7$N, the elastic cross section was calculated for the first time using the SCOP method.
% Additionally, the total ionization and excitation cross sections were obtained from~\cite{sinha2020}, and the attachment cross section from~\cite{chachereau2018}.
% The authors studied the dielectric strength of C$_4$F$_7$N-CO$_2$ mixtures under various pressures and gas mixture ratios.
% They observed that the simulation results had good agreement with experimental observations regarding the breakdown threshold.

The main novelty of our work is that we for the first time use 3D particle simulations, with which we can capture the stochastic inception and branching of streamers in C$_4$F$_7$N-CO$_2$ mixtures.
These stochastic aspects are particularly important since these mixtures appear to lack an effective photoionization mechanism, which results in much more irregular discharge growth than for example in air.
% Although our simulations always exhibit streamer branching, it would be interesting to explore the possibility of generating single streamers in these mixtures in future work.

Another difference is that we have included a stochastic background ionization process in some of the simulations, which produces electron-ion pairs at a certain rate over time.
In previous simulations of positive discharges, an initial background ionization density was assumed.
We observed that an initial electron density would quickly decay due to rapid attachment (unless the background field $E_\mathrm{bg}$ was above the critical field $E_\mathrm{k}$), after which positive streamer propagation was hardly possible.
However, both approaches are rather artificial, and further work is necessary to better understand free electron sources in such mixtures, such as photoionization and electron detachment from negative ions.

Our observation that streamers in C$_4$F$_7$N-CO$_2$ mixtures can only keep propagating if the background electric field $E_\mathrm{bg}$ is approximately the critical electric field $E_\mathrm{k}$ (due to rapid electron attachment) is in agreement with previous work.
However, the observed discharge growth in our 3D simulations differs significantly from that observed in previous 2D fluid simulations.
First, we observe stochastic growth with frequent branching, both for negative and positive polarities.
Second, we observe that new negative streamers can form behind previous ones, forming a chain of negative streamers.
Third, in 2D Cartesian simulations we see that positive discharges extend due to incoming negative streamers that connect to existing channels.

\section{Conclusions and outlook}\label{sec:con-and-outlook}

We have simulated negative and positive streamers in C$_4$F$_7$N-CO$_2$ mixtures at 300\,K and 1\,bar, using a 3D and 2D particle-in-cell model.
Negative streamers were able to start from a small number of initial electrons near an electrode, whereas positive streamers required a sustained source of free electrons ahead of them for their propagation.
We included an artificial stochastic background ionization process to provide such free electrons.

For negative streamers, a short conductive channel was observed behind the streamer head due to the fast attachment of electrons to C$_4$F$_7$N.
Due to the short channel, these discharges did not gain as much field enhancement as streamers in most other gases, so the background field required for streamer propagation was approximately the critical electric field.
Surprisingly, new negative streamers could be generated behind the previous ones when a stochastic background ionization process was included.
In this manner a chain of negative streamers could form.

Simulating positive streamers required the inclusion of the artificial background ionization process, and it was much more challenging due to the presence of extremely high fields and electron densities.
In 3D, we were only able to simulate the early inception stage of positive discharges.
To qualitatively investigate the growth of these discharges, we performed 2D Cartesian simulations.
We observed that positive discharge growth resulted from incoming negative streamers that connected to existing channels.
This led to highly irregular discharge growth, in which branching was determined by the locations of free electrons ahead of the discharge.

If there is indeed no effective photoionization mechanism in C$_4$F$_7$N-CO$_2$ mixtures, our results suggest that negative streamers will propagate more easily than positive ones.
This is in contrast to the behavior in air, where positive streamers initiate and propagate more easily than negative streamers~\cite{briels2008, starikovskiy2020}.

\textbf{Outlook.}
A better understanding of free electron sources such as photoionization and electron detachment is important for the modeling of streamer-like discharges in C$_4$F$_7$N-CO$_2$ mixtures.
Including ion kinetics could also be important, as was demonstrated in~\cite{hosl2019}.
% in which the authors showed that taking detachment of ions into account yields values at (55 ± 10)\% of the obtained rate coefficient of~\cite{chachereau2018}.
It would be valuable if streamer propagation could be captured experimentally, so that simulations and experiments could directly be compared.
Furthermore, a novel type of simulation model might be required to study positive discharges in these mixtures, because 3D simulations are currently too expensive due to the complex discharge structure and the extremely high electric fields and electron densities.

Finally, it would be interesting to consider additional ionization mechanisms.
First, future research could include secondary electron emission due to ions to more realistically study the inception of negative streamers, as in e.g.~\cite{babaeva2016a}.
Second, when the fraction of C$_4$F$_7$N or the pressure is increased together with the applied electric field, it could be interesting to include field ionization as in e.g.~\cite{abbas2023}.

% \subsubsection{Input data}

% There appears some failed runs when using Bolsig+ to calculate transport data for 4\% and 5\% C$_4$F$_7$N.
% new: failed at 4\%, 15\% and 20\%.

% \subsubsection{TODO}
% When simulating streamers in new gases, such as C$_4$F$_7$N-CO$_2$ mixtures, the so-called verification and validation is required~\cite{bagheri2018, li2021a}.
% Due to the limited availability of experimental data on streamers in C$_4$F$_7$N-CO$_2$ mixtures, here we focus on verification by comparing our simulations with past computational work and discussing their agreements or discrepancies.

\ack
B.G. was funded by the China Scholarship Council (CSC) (Grant No.\,201906280436).
We thank Prof. Dr. Christian Franck from Swiss Federal Institute of Technology and Prof. Dr. Bobby Antony from Indian Institute of Technology for sharing cross section data for C$_4$F$_7$N, and we thank Dr. Bastiaan Braams at CWI for valuable discussions.

\section*{Data availability statement}
The data that support the findings of this study are openly available at the following URL/DOI: \url{https://doi.org/10.5281/zenodo.8214598}.

\appendix

\section{Effect of the color scale on streamer visualization}\label{sec:neg-effect-of-color-scale}

\begin{figure*}
    \centering
    \includegraphics[width=1\textwidth]{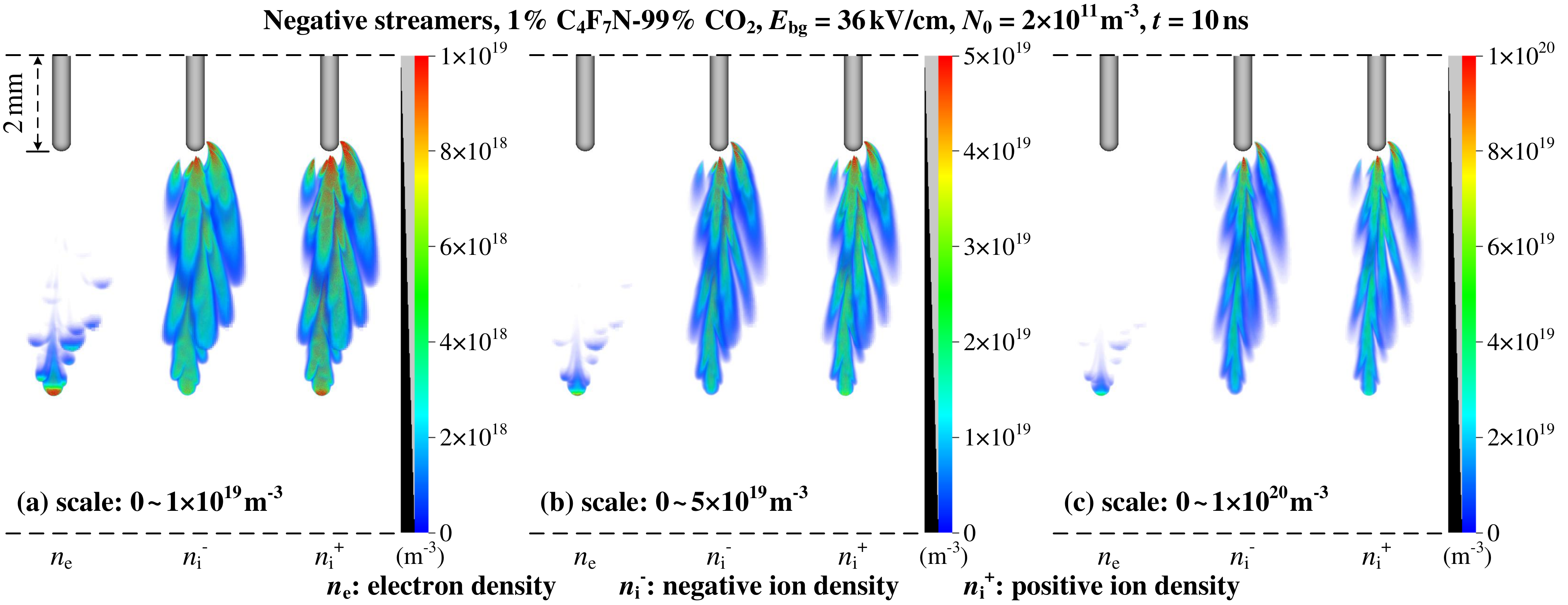}
    \caption{The case of negative streamers in figure~\ref{fig:neg_different_E}(b) at $t=10$\,ns.
    Shown is the electron density $n_\mathrm{e}$, negative ion density $n_\mathrm{i}^{-}$ and positive ion density $n_\mathrm{i}^{+}$ through Visit's 3D volume rendering with different color scales.
    With a higher maximum value of the color scale streamer channels appear to be thinner.}
    \label{fig:neg_different_scales}
\end{figure*}

Figure~\ref{fig:neg_different_scales} presents results with different color scales for negative streamers shown in figure~\ref{fig:neg_different_E}(b) at $t=10$\,ns, with the maximum scale value being $1\times10^{19}\,\mathrm{m}^{-3}$, $5\times10^{19}\,\mathrm{m}^{-3}$ and $1\times10^{20}\,\mathrm{m}^{-3}$, respectively.
With a higher value of the maximum, the streamers appear to be thinner.
% For optimal visualization, suitable color scales are therefore required. 

Throughout the paper, a linear scale ranging from 0 to $1\times10^{19}\,\mathrm{m}^{-3}$ is used for the electron density $n_\mathrm{e}$ and a linear scale ranging from 0 to $1\times10^{20}\,\mathrm{m}^{-3}$ is used for the positive ion density $n_\mathrm{i}^{+}$, except for the 3D simulations of positive discharges shown in section~\ref{sec:pos-3d-results}, which have much higher densities, see figures~\ref{fig:pos_3d_different_bir} and~\ref{fig:pos_3d_cross_sections}.

\section{Computational costs}\label{sec:computational-costs}

The primary drawback of a 3D PIC-MCC model is its high computational cost, due to the large number of simulation particles required.
The simulations in this paper were performed on workstations with 8 cores and 32--64\,GB of RAM, using at most $5 \times 10^8$ simulation particles.
The run time for the negative cases shown in section~\ref{sec:neg-streamers} was about 1 week, whereas the 3D simulations of positive streamer inception shown in section~\ref{sec:pos-3d-results}  took about 2 weeks.
The 2D simulations of positive streamers shown in section~\ref{sec:pos-2d-results} only took a few hours.

% Computational cost:
% We ran the codes on a supercomputer with 4 AMD Opteron 6344 12-core processors and 120 GB of RAM was used.
% jannis: The simulations were performed on a single node containing two Xeon E5-2680v4 processors (2 × 14 cores, at 2.4 GHz). The simulations ran for up to 24 hours, using up to 1e8 grid cells.
% jannis: The simulations presented in section 3 were performed on nodes with 12–16 cores. They took up to 48 h, using at most 6e7 simulation particles.
% zhen: The 3D particle simulations were performed on Cartesius, the Dutch national supercomputer. A single thin node with an Intel Xeon E5-2690 v3 (Haswell) 24-core processor and 64 GB of RAM was used. Computations ran for up to five days, with up to 600 million particles.

\section*{References}
\normalem
\bibliography{references}

\begin{thebibliography}{10}

\bibitem{christophorou1997}
L.G. Christophorou, J.K. Olthoff, and R.J. Van~Brunt.
\newblock Sulfur hexafluoride and the electric power industry.
\newblock {\em IEEE Electrical Insulation Magazine}, 13(5):20--24, September
  1997.

\bibitem{malik1978}
N.~H. Malik and A.~H. Qureshi.
\newblock Breakdown {{Mechanisms}} in {{Sulphur-Hexafluoride}}.
\newblock {\em IEEE Transactions on Electrical Insulation}, EI-13(3):135--145,
  June 1978.

\bibitem{boggs1990}
S.A. Boggs and H.-H. Schramm.
\newblock Current interruption and switching in sulphur hexafluoride.
\newblock {\em IEEE Electrical Insulation Magazine}, 6(1):12--17, January 1990.

\bibitem{ray2017}
Eric~A. Ray, Fred~L. Moore, James~W. Elkins, Karen~H. Rosenlof, Johannes~C.
  Laube, Thomas R{\"o}ckmann, Daniel~R. Marsh, and Arlyn~E. Andrews.
\newblock Quantification of the {{SF}}{\textsubscript{6}} lifetime based on
  mesospheric loss measured in the stratospheric polar vortex.
\newblock {\em Journal of Geophysical Research: Atmospheres},
  122(8):4626--4638, 2017.

\bibitem{IPCC2021}
V.~Masson-Delmotte, P.~Zhai, A.~Pirani, S.L. Connors, C.~Péan, S.~Berger,
  N.~Caud, Y.~Chen, L.~Goldfarb, M.I. Gomis, M.~Huang, K.~Leitzell, E.~Lonnoy,
  J.B.R. Matthews, T.K. Maycock, T.~Waterfield, O.~Yelekçi, R.~Yu, and
  B.~Zhou~(eds.).
\newblock {{IPCC}}, 2021: Climate change 2021: The physical science basis.
  contribution of working group {{I}} to the sixth assessment report of the
  intergovernmental panel on climate change.
\newblock Technical report, Cambridge University Press, Cambridge, United
  Kingdom and New York, NY, USA, 2021.

\bibitem{kieffel2016}
Yannick Kieffel, Todd Irwin, Philippe Ponchon, and John Owens.
\newblock Green {{Gas}} to replace {{SF}}{\textsubscript{6}} in {{Electrical
  Grids}}.
\newblock {\em IEEE Power and Energy Magazine}, 14(2):32--39, March 2016.

\bibitem{rabie2018}
Mohamed Rabie and Christian~M. Franck.
\newblock Assessment of {{Eco-friendly Gases}} for {{Electrical Insulation}} to
  {{Replace}} the most potent industrial greenhouse gas
  {{SF}}{\textsubscript{6}}.
\newblock {\em Environmental Science \& Technology}, 52(2):369--380, January
  2018.

\bibitem{nechmi2016}
H.~E. Nechmi, A.~Beroual, A.~Girodet, and P.~Vinson.
\newblock Fluoronitriles/{{CO}}{\textsubscript{2}} gas mixture as promising
  substitute to {{SF}}{\textsubscript{6}} for insulation in high voltage
  applications.
\newblock {\em IEEE Transactions on Dielectrics and Electrical Insulation},
  23(5):2587--2593, October 2016.

\bibitem{zhang2017a}
Xiaoxing Zhang, Yi~Li, Song Xiao, Shuangshuang Tian, Zaitao Deng, and Ju~Tang.
\newblock Theoretical study of the decomposition mechanism of environmentally
  friendly insulating medium
  {{C}}{\textsubscript{3}}{{F}}{\textsubscript{7}}{{CN}} in the presence of
  {{H}}{\textsubscript{2}}{{O}} in a discharge.
\newblock {\em Journal of Physics D: Applied Physics}, 50(32):325201, August
  2017.

\bibitem{fu2019}
Yuwei Fu, Aijun Yang, Xiaohua Wang, and Mingzhe Rong.
\newblock Theoretical study of the decomposition mechanism of
  {{C}}{\textsubscript{4}}{{F}}{\textsubscript{7}}{{N}}.
\newblock {\em Journal of Physics D: Applied Physics}, 52(24):245203, June
  2019.

\bibitem{chen2019}
Li~Chen, Boya Zhang, Jiayu Xiong, Xingwen Li, and Anthony~B. Murphy.
\newblock Decomposition mechanism and kinetics of iso-{{C4}} perfluoronitrile
  ({{C}}{\textsubscript{4}}{{F}}{\textsubscript{7}}{{N}}) plasmas.
\newblock {\em Journal of Applied Physics}, 126(16):163303, October 2019.

\bibitem{chen2020a}
Li~Chen, Boya Zhang, Tao Yang, Yunkun Deng, Xingwen Li, and Anthony~B Murphy.
\newblock Thermal decomposition characteristics and kinetic analysis of
  {{C}}{\textsubscript{4}}{{F}}{\textsubscript{7}}{{N}}/{{CO}}{\textsubscript{2}}
  gas mixture.
\newblock {\em Journal of Physics D: Applied Physics}, 53(5):055502, January
  2020.

\bibitem{li2020d}
Yi~Li, Xiaoxing Zhang, Ji~Zhang, Cheng Xie, Xianjun Shao, Ziling Wang, Dachang
  Chen, and Song Xiao.
\newblock Study on the thermal decomposition characteristics of
  {{C}}{\textsubscript{4}}{{F}}{\textsubscript{7}}{{N}}\textendash{{CO}}{\textsubscript{2}}
  mixture as eco-friendly gas-insulating medium.
\newblock {\em High Voltage}, 5(1):46--52, 2020.

\bibitem{xiao2021a}
Song Xiao, ShengYao Shi, Yi~Li, Fanchao Ye, Yalong Li, Shuangshuang Tian,
  Ju~Tang, and Xiaoxing Zhang.
\newblock Review of decomposition characteristics of eco-friendly gas
  insulating medium for high-voltage gas-insulated equipment.
\newblock {\em Journal of Physics D: Applied Physics}, 54(37):373002, September
  2021.

\bibitem{nechmi2017}
H.E. Nechmi, A.~Beroual, A.~Girodet, and P.~Vinson.
\newblock Effective ionization coefficients and limiting field strength of
  fluoronitriles-{{CO}}{\textsubscript{2}} mixtures.
\newblock {\em IEEE Transactions on Dielectrics and Electrical Insulation},
  24(2):886--892, April 2017.

\bibitem{chachereau2018}
A~Chachereau, A~H{\"o}sl, and C~M Franck.
\newblock Electrical insulation properties of the perfluoronitrile
  {{C}}{\textsubscript{4}}{{F}}{\textsubscript{7}}{{N}}.
\newblock {\em Journal of Physics D: Applied Physics}, 51(49):495201, December
  2018.

\bibitem{qin2019}
Zhaoyu Qin, Yunxiang Long, Zhenyu Shen, Cheng Chen, Liping Guo, and Wenjun
  Zhou.
\newblock Ionization and {{Attachment Coefficients}} in
  {{C}}{\textsubscript{4}}{{F}}{\textsubscript{7}}{{N}} {{Gas Measured}} by the
  {{Steady-State Townsend Method}}.
\newblock {\em Applied Sciences}, 9(18):3686, September 2019.

\bibitem{yi2020}
Chengqian Yi, Zhikang Yuan, Youping Tu, Ying Zhang, and Cong Wang.
\newblock Measurements of discharge parameters in
  {{C}}{\textsubscript{3}}{{F}}{\textsubscript{7}}{{CN}}/{{CO}}{\textsubscript{2}}
  and
  {{C}}{\textsubscript{3}}{{F}}{\textsubscript{7}}{{CN}}/{{N}}{\textsubscript{2}}
  gas mixtures by {{SST}}.
\newblock {\em IEEE Transactions on Dielectrics and Electrical Insulation},
  27(3):1015--1021, June 2020.

\bibitem{long2020}
Yunxiang Long, Liping Guo, Cheng Chen, Zhenyu Shen, Yiheng Chen, Fang Li, and
  Wenjun Zhou.
\newblock Measurement of {{Ionization}} and {{Attachment Coefficients}} in
  {{C}}{\textsubscript{4}}{{F}}{\textsubscript{7}}{{N}}/{{CO}}{\textsubscript{2}}
  {{Gas Mixture}} as {{Substitute Gas}} to {{SF}}{\textsubscript{6}}.
\newblock {\em IEEE Access}, 8:76790--76795, 2020.

\bibitem{zhang2022}
Boya Zhang, Jiayu Xiong, Mai Hao, Yuyang Yao, Xingwen Li, and Anthony~B.
  Murphy.
\newblock Pulsed {{Townsend}} measurement of electron swarm parameters in
  {{C}}{\textsubscript{4}}{{F}}{\textsubscript{7}}{{N}}\textendash
  {{CO}}{\textsubscript{2}} and
  {{C}}{\textsubscript{4}}{{F}}{\textsubscript{7}}{{N}}\textendash
  {{N}}{\textsubscript{2}} mixtures as eco-friendly insulation gas.
\newblock {\em Journal of Applied Physics}, 131(3):033304, January 2022.

\bibitem{tu2018}
Youping Tu, Yi~Cheng, Cong Wang, Xin Ai, Fuwen Zhou, and Geng Chen.
\newblock Insulation characteristics of
  fluoronitriles/{{CO}}{\textsubscript{2}} gas mixture under {{DC}} electric
  field.
\newblock {\em IEEE Transactions on Dielectrics and Electrical Insulation},
  25(4):1324--1331, August 2018.

\bibitem{zhang2018}
Boya Zhang, Nenad Uzelac, and Yang Cao.
\newblock Fluoronitrile/{{CO}}{\textsubscript{2}} mixture as an eco-friendly
  alternative to {{SF}}{\textsubscript{6}} for medium voltage switchgears.
\newblock {\em IEEE Transactions on Dielectrics and Electrical Insulation},
  25(4):1340--1350, August 2018.

\bibitem{zhao2018}
Hu~Zhao, Xingwen Li, Nian Tang, Xu~Jiang, Ze~Guo, and Hui Lin.
\newblock Dielectric properties of fluoronitriles/{{CO}}{\textsubscript{2}} and
  {{SF}}{\textsubscript{6}}/{{N}}{\textsubscript{2}} mixtures as a possible
  {{SF}}{\textsubscript{6}}-{{substitute}} gas.
\newblock {\em IEEE Transactions on Dielectrics and Electrical Insulation},
  25(4):1332--1339, August 2018.

\bibitem{zhang2019a}
Xiaoxing Zhang, Qi~Chen, Ji~Zhang, Yi~Li, Song Xiao, Ran Zhuo, and Ju~Tang.
\newblock Experimental {{Study}} on {{Power Frequency Breakdown
  Characteristics}} of
  {{C}}{\textsubscript{4}}{{F}}{\textsubscript{7}}{{N}}/{{CO}}{\textsubscript{2}}
  {{Gas Mixture Under Quasi-Homogeneous Electric Field}}.
\newblock {\em IEEE Access}, 7:19100--19108, 2019.

\bibitem{nechmi2020}
Houssem~Eddine Nechmi, Mohammed El~Amine Slama, Abderrahmane~(Manu) Haddad, and
  Gordon Wilson.
\newblock {{AC Volume Breakdown}} and {{Surface Flashover}} of a 4\%
  {{Novec}}{$^\textrm{TM}$} 4710/96\% {{CO}}{\textsubscript{2}} {{Gas Mixture
  Compared}} to {{CO}}{\textsubscript{2}} in {{Highly Nonhomogeneous Fields}}.
\newblock {\em Energies}, 13(7):1710, April 2020.

\bibitem{zhang2020b}
Tianran Zhang, Wenjun Zhou, Jianhui Yu, and Zhuangxin Yu.
\newblock Insulation properties of
  {{C}}{\textsubscript{4}}{{F}}{\textsubscript{7}}{{N}}/{{CO}}{\textsubscript{2}}
  mixtures under lightning impulse.
\newblock {\em IEEE Transactions on Dielectrics and Electrical Insulation},
  27(1):181--188, February 2020.

\bibitem{zhou2020}
Wenjun Zhou, Tianran Zhang, and Lingzhi Wang.
\newblock Breakdown {{Characteristics}} of
  {{C}}{\textsubscript{4}}{{F}}{\textsubscript{7}}{{N}}/{{CO}}{\textsubscript{2}}
  {{Mixtures Under Steep-Wavefront Impulse Voltages}}.
\newblock {\em IEEE Access}, 8:29291--29298, 2020.

\bibitem{nechmi2021}
Houssem~Eddine Nechmi, Michail Michelarakis, Abderrahmane (Manu)~Haddad, and
  Gordon Wilson.
\newblock Clarifications on the {{Behavior}} of {{Alternative Gases}} to
  {{SF}}{\textsubscript{6}} in {{Divergent Electric Field Distributions}} under
  {{AC Voltage}}.
\newblock {\em Energies}, 14(4):1065, February 2021.

\bibitem{bahdad2022}
Faisal~Omar Bahdad, Lujia Chen, Prem Ranjan, and Simon Rowland.
\newblock Effects of {{DC Polarity}} and {{Field Uniformity}} on {{Breakdown}}
  of {{SF}}{\textsubscript{6}} and
  {{C}}{\textsubscript{3}}{{F}}{\textsubscript{7}}{{CN}}/{{CO}}{\textsubscript{2}}
  mixture.
\newblock {\em IEEE Transactions on Dielectrics and Electrical Insulation},
  29(6):2227--2235, December 2022.

\bibitem{lin2022}
Xin Lin, Jia Zhang, Jianyuan Xu, Jianying Zhong, Yu~Song, and Youpeng Zhang.
\newblock Dynamic {{Dielectric Strength}} of
  {{C}}{\textsubscript{3}}{{F}}{\textsubscript{7}}{{CN}}/{{CO}}{\textsubscript{2}}
  and
  {{C}}{\textsubscript{3}}{{F}}{\textsubscript{7}}{{CN}}/{{N}}{\textsubscript{2}}
  gas mixtures in {{High Voltage Circuit Breakers}}.
\newblock {\em IEEE Transactions on Power Delivery}, 37(5):4032--4041, October
  2022.

\bibitem{zheng2019}
Yu~Zheng, Xianglian Yan, Weijiang Chen, Wenjun Zhou, Shizhuo Hu, and Han Li.
\newblock Calculation of electrical insulation of
  {{C}}{\textsubscript{4}}{{F}}{\textsubscript{7}}{{N}}/{{CO}}{\textsubscript{2}}
  mixed gas by avalanche characteristics of pure gas.
\newblock {\em Plasma Research Express}, 1(2):025013, June 2019.

\bibitem{zhang2020a}
Boya Zhang, Li~Chen, Xingwen Li, Ze~Guo, Yunjie Pu, and Nian Tang.
\newblock Evaluating the dielectric strength of promising
  {{SF}}{\textsubscript{6}} alternatives by {{DFT}} calculations and {{DC}}
  breakdown tests.
\newblock {\em IEEE Transactions on Dielectrics and Electrical Insulation},
  27(4):1187--1194, August 2020.

\bibitem{zheng2020}
Yu~Zheng, Wenjun Zhou, Han Li, and Chenxi Ma.
\newblock Experimental and calculation study on insulation strength of
  {{C}}{\textsubscript{4}}{{F}}{\textsubscript{7}}{{N}}/{{CO}}{\textsubscript{2}}
  at low temperature.
\newblock {\em IEEE Transactions on Dielectrics and Electrical Insulation},
  27(4):1102--1109, August 2020.

\bibitem{nijdam2020a}
Sander Nijdam, Jannis Teunissen, and Ute Ebert.
\newblock The physics of streamer discharge phenomena.
\newblock {\em Plasma Sources Science and Technology}, 29(10):103001, November
  2020.

\bibitem{vu-cong2020}
T.~{Vu-Cong}, C.~Toigo, G.~Ortiz, M.~Dalstein, F.~Jacquier, and A.~Girodet.
\newblock Numerical simulation of partial discharge current pulse:
  {{Comparison}} between {{SF}}{\textsubscript{6}},
  {{Fluoronitrile}}\textendash {{CO}}{\textsubscript{2}} mixture and
  {{Fluoroketone}}\textendash {{CO}}{\textsubscript{2}} mixture.
\newblock In {\em 2020 {{IEEE Conference}} on {{Electrical Insulation}} and
  {{Dielectric Phenomena}} ({{CEIDP}})}, pages 403--406, October 2020.

\bibitem{wang2021a}
Feng Wang, Lanbo Wang, She Chen, Qiuqin Sun, Lipeng Zhong, and Chijie Zhuang.
\newblock Numerical {{Simulation}} of the {{Discharge Dynamics}} of
  {{C}}{$_{4}$}{{F}}{$_{7}$}{{N-N}}{$_2$} and the {{Influence}} of {{Buffer
  Gas}}.
\newblock {\em IEEE Transactions on Plasma Science}, 49(7):2048--2054, July
  2021.

\bibitem{fan2022}
Binhai Fan, Xiaoli Zhou, Yong Qian, and Yiming Zang.
\newblock Simulation of positive surface discharge in
  {{C}}{$_{4}$}{{F}}{$_{7}$}{{N}} gas mixture.
\newblock In {\em 2022 7th {{Asia Conference}} on {{Power}} and {{Electrical
  Engineering}} ({{ACPEE}})}, pages 1604--1608, April 2022.

\bibitem{gao2022}
Qingqing Gao, Xiaohua Wang, Kazimierz Adamiak, Xiangcheng Qi, Aijun Yang,
  Dingxin Liu, Chunping Niu, and Jiawei Zhang.
\newblock Negative corona discharge mechanism in
  {{C}}{\textsubscript{4}}{{F}}{\textsubscript{7}}{{N}}\textendash{{CO}}{\textsubscript{2}}
  and
  {{C}}{\textsubscript{4}}{{F}}{\textsubscript{7}}{{N}}\textendash{{N}}{\textsubscript{2}}
  mixtures.
\newblock {\em AIP Advances}, 12(9):095101, September 2022.

\bibitem{yan2023}
Xinfeng Yan, Xiaoli Zhou, Ze~Li, Yong Qian, and Gehao Sheng.
\newblock Numerical simulation of streamer discharge with different electrode
  shapes in {{C}}{\textsubscript{4}}{{F}}{\textsubscript{7}}{{N}}.
\newblock {\em AIP Advances}, 13(3):035238, March 2023.

\bibitem{yan2023a}
Xinfeng Yan, Xiaoli Zhou, Ze~Li, Yong Qian, and Gehao Sheng.
\newblock Surface {{Discharge Characteristics}} and {{Numerical Simulation}} in
  {{C}}{\textsubscript{4}}{{F}}{\textsubscript{7}}{{N}}/{{CO}}{\textsubscript{2}}
  {{Mixture}}.
\newblock {\em Applied Sciences}, 13(3):1409, January 2023.

\bibitem{petrovic2009}
Z~Lj Petrovi{\'c}, S~Dujko, D~Mari{\'c}, G~Malovi{\'c}, {\v Z}~Nikitovi{\'c},
  O~{\v S}a{\v s}i{\'c}, J~Jovanovi{\'c}, V~Stojanovi{\'c}, and
  M~{Radmilovi{\'c}-Ra{\dj}enovi{\'c}}.
\newblock Measurement and interpretation of swarm parameters and their
  application in plasma modelling.
\newblock {\em Journal of Physics D: Applied Physics}, 42(19):194002, October
  2009.

\bibitem{zhang2020}
Boya Zhang, Jiayu Xiong, Li~Chen, Xingwen Li, and Anthony~B Murphy.
\newblock Fundamental physicochemical properties of
  {{SF}}{\textsubscript{6}}-alternative gases: A review of recent progress.
\newblock {\em Journal of Physics D: Applied Physics}, 53(17):173001, April
  2020.

\bibitem{xiong2017}
Jiayu Xiong, Xingwen Li, Jian Wu, Xiaoxue Guo, and Hu~Zhao.
\newblock Calculations of total electron-impact ionization cross sections for
  {{Fluoroketone C}}{\textsubscript{5}}{{F}}{\textsubscript{10}}{{O}} and
  {{Fluoronitrile C}}{\textsubscript{4}}{{F}}{\textsubscript{7}}{{N}} using
  modified {{Deutsch}}\textendash{{M\"ark}} formula.
\newblock {\em Journal of Physics D: Applied Physics}, 50(44):445206, November
  2017.

\bibitem{wang2019a}
Feng Wang, Qiaowen Dun, She Chen, Lipeng Zhong, Xiaopeng Fan, and Li~Li.
\newblock Calculations of total electron impact ionization cross sections for
  fluoroketone and fluoronitrile.
\newblock {\em IEEE Transactions on Dielectrics and Electrical Insulation},
  26(5):1693--1700, October 2019.

\bibitem{zhong2019a}
Linlin Zhong, Jie Xu, Xiaohua Wang, and Mingzhe Rong.
\newblock Electron-impact ionization cross sections of new
  {{SF}}{\textsubscript{6}} replacements: {{A}} method of combining
  {{Binary-Encounter-Bethe}} ({{BEB}}) and {{Deutsch-M\"ark}} ({{DM}})
  formalism.
\newblock {\em Journal of Applied Physics}, 126(19):193302, November 2019.

\bibitem{rankovic2019}
M.~Rankovi{\'c}, J.~Chalabala, M.~Zawadzki, J.~Ko{\v c}i{\v s}ek,
  P.~Slav{\'i}{\v c}ek, and J.~Fedor.
\newblock Dissociative ionization dynamics of dielectric gas
  {{C}}{\textsubscript{3}}{{F}}{\textsubscript{7}}{{CN}}.
\newblock {\em Physical Chemistry Chemical Physics}, 21(30):16451--16458, 2019.

\bibitem{rankovic2020}
M.~Rankovi{\'c}, Ragesh Kumar T~P, P.~Nag, J.~Ko{\v c}i{\v s}ek, and J.~Fedor.
\newblock Temporary anions of the dielectric gas
  {{C}}{\textsubscript{3}}{{F}}{\textsubscript{7}}{{CN}} and their decay
  channels.
\newblock {\em The Journal of Chemical Physics}, 152(24):244304, June 2020.

\bibitem{sinha2020}
Nidhi Sinha, Vraj~Manishkumar Patel, and Bobby Antony.
\newblock Ionization cross sections for plasma relevant molecules.
\newblock {\em Journal of Physics B: Atomic, Molecular and Optical Physics},
  53(14):145101, July 2020.

\bibitem{zhang2022b}
Jianwei Zhang, Nidhi Sinha, Ming Jiang, Hongguang Wang, Yongdong Li, Bobby
  Antony, and Chunliang Liu.
\newblock {{DC Breakdown Characteristics}} of
  {{C}}{$_{4}$}{{F}}{$_{7}$}{{N}}/{{CO}}{$_2$} {{Mixtures With Particle-in-Cell
  Simulation}}.
\newblock {\em IEEE Transactions on Dielectrics and Electrical Insulation},
  29(3):1005--1010, June 2022.

\bibitem{zhang2023}
Boya Zhang, Mai Hao, Yuyang Yao, Jiayu Xiong, Xingwen Li, Anthony~B Murphy,
  Nidhi Sinha, Bobby Antony, and Harindranath~B Ambalampitiya.
\newblock Determination and assessment of a complete and self-consistent
  electron-neutral collision cross-section set for the
  {{C}}{\textsubscript{4}}{{F}}{\textsubscript{7}}{{N}} molecule.
\newblock {\em Journal of Physics D: Applied Physics}, 56(13):134001, March
  2023.

\bibitem{hosl2019}
Andreas H{\"o}sl, Alise Chachereau, Juriy Pachin, and Christian~M Franck.
\newblock Identification of the discharge kinetics in the perfluoro-nitrile
  {{C}}{\textsubscript{4}}{{F}}{\textsubscript{7}}{{N}} with swarm and
  breakdown experiments.
\newblock {\em Journal of Physics D: Applied Physics}, 52(23):235201, June
  2019.

\bibitem{teunissen2016}
Jannis Teunissen and Ute Ebert.
\newblock {{3D PIC-MCC}} simulations of discharge inception around a sharp
  anode in nitrogen/oxygen mixtures.
\newblock {\em Plasma Sources Science and Technology}, 25(4):044005, August
  2016.

\bibitem{wang2022}
Zhen Wang, Anbang Sun, and Jannis Teunissen.
\newblock A comparison of particle and fluid models for positive streamer
  discharges in air.
\newblock {\em Plasma Sources Science and Technology}, 31(1):015012, January
  2022.

\bibitem{verlet1967}
Loup Verlet.
\newblock Computer ``{{Experiments}}'' on {{Classical Fluids}}. {{I}}.
  {{Thermodynamical Properties}} of {{Lennard-Jones Molecules}}{$^*$}.
\newblock {\em Physical Review}, 159(1):98--103, July 1967.

\bibitem{koura1986}
Katsuhisa Koura.
\newblock Null-collision technique in the direct-simulation {{Monte Carlo}}
  method.
\newblock {\em Physics of Fluids}, 29(11):3509, 1986.

\bibitem{teunissen2014a}
Jannis Teunissen and Ute Ebert.
\newblock Controlling the weights of simulation particles: Adaptive particle
  management using k-d trees.
\newblock {\em Journal of Computational Physics}, 259:318--330, February 2014.

\bibitem{teunissen2018}
Jannis Teunissen and Ute Ebert.
\newblock Afivo: {{A}} framework for quadtree/octree {{AMR}} with shared-memory
  parallelization and geometric multigrid methods.
\newblock {\em Computer Physics Communications}, 233:156--166, December 2018.

\bibitem{teunissen2023}
Jannis Teunissen and Francesca Schiavello.
\newblock Geometric multigrid method for solving {{Poisson}}'s equation on
  octree grids with irregular boundaries.
\newblock {\em Computer Physics Communications}, 286:108665, May 2023.

\bibitem{XJTUAETLab_database}
XJTUAETLab database ({C}\textsubscript{4}{F}\textsubscript{7}N)
  \url{www.lxcat.net} (retrieved 26 April 2023).

\bibitem{IST-Lisbon_database}
IST-Lisbon database ({CO}\textsubscript{2}) \url{www.lxcat.net} (retrieved 26
  April 2023).

\bibitem{grofulovic2016}
Marija Grofulovi{\'c}, Lu{\'i}s~L Alves, and Vasco Guerra.
\newblock Electron-neutral scattering cross sections for
  {{CO}}{\textsubscript{2}}: A complete and consistent set and an assessment of
  dissociation.
\newblock {\em Journal of Physics D: Applied Physics}, 49(39):395207, October
  2016.

\bibitem{eco_phelps_database}
Phelps database ({N}\textsubscript{2}, {O}\textsubscript{2})
  \url{www.lxcat.net} (retrieved 26 April 2023).

\bibitem{hagelaar2005}
G~J~M Hagelaar and L~C Pitchford.
\newblock Solving the {{Boltzmann}} equation to obtain electron transport
  coefficients and rate coefficients for fluid models.
\newblock {\em Plasma Sources Science and Technology}, 14(4):722--733, November
  2005.

\bibitem{seeger2017}
M~Seeger, J~Avaheden, S~Pancheshnyi, and T~Votteler.
\newblock Streamer parameters and breakdown in {{CO}}{\textsubscript{2}}.
\newblock {\em Journal of Physics D: Applied Physics}, 50(1):015207, January
  2017.

\bibitem{seeger2018}
Martin Seeger, Torsten Votteler, Jonas Ekeberg, Sergey Pancheshnyi, and Luis
  S{\'a}nchez.
\newblock Streamer and leader breakdown in air at atmospheric pressure in
  strongly non-uniform fields in gaps less than one metre.
\newblock {\em IEEE Transactions on Dielectrics and Electrical Insulation},
  25(6):2147--2156, December 2018.

\bibitem{pancheshnyi2014}
Sergey Pancheshnyi.
\newblock Photoionization produced by low-current discharges in
  {{O}}{\textsubscript{2}}, air, {{N}}{\textsubscript{2}} and
  {{CO}}{\textsubscript{2}}.
\newblock {\em Plasma Sources Science and Technology}, 24(1):015023, December
  2014.

\bibitem{nijdam2011}
S~Nijdam, G~Wormeester, E~M {van Veldhuizen}, and U~Ebert.
\newblock Probing background ionization: Positive streamers with varying pulse
  repetition rate and with a radioactive admixture.
\newblock {\em Journal of Physics D: Applied Physics}, 44(45):455201, November
  2011.

\bibitem{HPV:VisIt}
Hank Childs, Eric Brugger, Brad Whitlock, Jeremy Meredith, Sean Ahern, David
  Pugmire, Kathleen Biagas, Mark Miller, Cyrus Harrison, Gunther~H. Weber, Hari
  Krishnan, Thomas Fogal, Allen Sanderson, Christoph Garth, E.~Wes Bethel,
  David Camp, Oliver R\"{u}bel, Marc Durant, Jean~M. Favre, and Paul
  Navr\'{a}til.
\newblock Vis{{I}}t: An end-user tool for visualizing and analyzing very large
  data.
\newblock In {\em High Performance Visualization--Enabling Extreme-Scale
  Scientific Insight}, pages 357--372. Chapman and Hall/CRC, October 2012.

\bibitem{guo2022d}
Baohong Guo, Xiaoran Li, Ute Ebert, and Jannis Teunissen.
\newblock A computational study of accelerating, steady and fading negative
  streamers in ambient air.
\newblock {\em Plasma Sources Science and Technology}, 31(9):095011, September
  2022.

\bibitem{bujotzek2015}
M~Bujotzek, M~Seeger, F~Schmidt, M~Koch, and C~Franck.
\newblock Experimental investigation of streamer radius and length in
  {{SF}}{\textsubscript{6}}.
\newblock {\em Journal of Physics D: Applied Physics}, 48(24):245201, June
  2015.

\bibitem{francisco2021a}
Hani Francisco, Behnaz Bagheri, and Ute Ebert.
\newblock Electrically isolated propagating streamer heads formed by strong
  electron attachment.
\newblock {\em Plasma Sources Science and Technology}, 30(2):025006, February
  2021.

\bibitem{allen1991}
N.L. Allen and M.~Boutlendj.
\newblock Study of the electric fields required for streamer propagation in
  humid air.
\newblock {\em IEE Proceedings A Science, Measurement and Technology},
  138(1):37, 1991.

\bibitem{li2022a}
Xiaoran Li, Baohong Guo, Anbang Sun, Ute Ebert, and Jannis Teunissen.
\newblock A computational study of steady and stagnating positive streamers in
  {{N}}{\textsubscript{2}}\textendash {{O}}{\textsubscript{2}} mixtures.
\newblock {\em Plasma Sources Science and Technology}, 31(6):065011, June 2022.

\bibitem{li2023}
Xiaoran Li, Anbang Sun, and Jannis Teunissen.
\newblock The effect of photoionization on positive streamers in
  {{CO}}{\textsubscript{2}} studied with {{2D}} particle-in-cell simulations.
\newblock (arXiv:2304.01531), April 2023.

\bibitem{nijdam2010}
S~Nijdam, F~M J~H {van de Wetering}, R~Blanc, E~M {van Veldhuizen}, and
  U~Ebert.
\newblock Probing photo-ionization: Experiments on positive streamers in pure
  gases and mixtures.
\newblock {\em Journal of Physics D: Applied Physics}, 43(14):145204, April
  2010.

\bibitem{teunissen2017}
Jannis Teunissen and Ute Ebert.
\newblock Simulating streamer discharges in {{3D}} with the parallel adaptive
  {{Afivo}} framework.
\newblock {\em Journal of Physics D: Applied Physics}, 50(47):474001, November
  2017.

\bibitem{long2019}
Yunxiang Long, Liping Guo, Zhenyu Shen, Cheng Chen, Yiheng Chen, Fang Li, and
  Wenjun Zhou.
\newblock Ionization and attachment coefficients in
  {{C}}{\textsubscript{4}}{{F}}{\textsubscript{7}}{{N}}/{{N}}{\textsubscript{2}}
  gas mixtures for use as a replacement to {{SF}}{\textsubscript{6}}.
\newblock {\em IEEE Transactions on Dielectrics and Electrical Insulation},
  26(4):1358--1362, August 2019.

\bibitem{briels2008}
T~M~P Briels, J~Kos, G~J~J Winands, E~M {van Veldhuizen}, and U~Ebert.
\newblock Positive and negative streamers in ambient air: Measuring diameter,
  velocity and dissipated energy.
\newblock {\em Journal of Physics D: Applied Physics}, 41(23):234004, December
  2008.

\bibitem{starikovskiy2020}
A~Yu Starikovskiy and N~L Aleksandrov.
\newblock How pulse polarity and photoionization control streamer discharge
  development in long air gaps.
\newblock {\em Plasma Sources Science and Technology}, 29(7):075004, July 2020.

\bibitem{babaeva2016a}
Natalia~Yu Babaeva, Dmitry~V Tereshonok, and George~V Naidis.
\newblock Fluid and hybrid modeling of nanosecond surface discharges: Effect of
  polarity and secondary electrons emission.
\newblock {\em Plasma Sources Science and Technology}, 25(4):044008, July 2016.

\bibitem{abbas2023}
Muhammad~Farasat Abbas, Yan~Liang He, Guang~Yu Sun, An~Bang Sun, Elsayed~Tag
  Eldin, and Sherif S.~M. Ghoneim.
\newblock Positive {{Streamer Initiation}} in
  {{SF}}{\textsubscript{6}}/{{CO}}{\textsubscript{2}} {{Based}} on {{Zener}}'s
  {{Field Ionization}}.
\newblock {\em IEEE Access}, 11:91767--91776, 2023.

\end{thebibliography}

\end{document}